\documentclass[]{aastex63}

\usepackage{graphicx}
\usepackage{epstopdf}
\usepackage{natbib}
\usepackage{float}
\graphicspath{{./}{figures/}}

\accepted{\today}
\submitjournal{ApJ}

\shorttitle{Stellar Rotation of T Tauri stars in the OSFC}
\shortauthors{Serna et al.}

\begin{document}

\title{Stellar Rotation of T Tauri stars in the Orion Star-Forming Complex.}
\correspondingauthor{Javier Serna}
\email{jserna@astro.unam.mx}

\author[0000-0001-7351-6540]{Javier Serna}
\affil{Instituto de Astronom\'{i}a, Universidad Aut\'{o}noma de M\'{e}xico,
Ensenada, B.C, M\'{e}xico}

\author[0000-0001-9797-5661]{Jesus Hernandez}
\affil{Instituto de Astronom\'{i}a, Universidad Aut\'{o}noma de M\'{e}xico,
Ensenada, B.C, M\'{e}xico}

\author[0000-0002-5365-1267]{Marina Kounkel}
\affil{Department of Physics and Astronomy, Vanderbilt University, VU Station 1807, Nashville, TN 37235, USA}
\affil{Department of Physics and Astronomy, Western Washington University, 516 High St, Bellingham, WA 98225}

\author[0000-0001-6647-862X]{Ezequiel Manzo-Mart\'{\i}nez}
\affil{Instituto de Astronom\'{i}a, Universidad Aut\'{o}noma de M\'{e}xico,
Ensenada, B.C, M\'{e}xico}
\affiliation{Mesoamerican Centre for Theoretical Physics, Universidad Aut\'onoma de Chiapas, Carretera Zapata Km. 4, Real del Bosque (Ter\'an), Tuxtla Guti\'errez 29040, Chiapas, M\'exico}

\author[0000-0002-1379-4204]{Alexandre Roman-Lopes}
\affiliation{Departamento de Astronom{\'i}a, Universidad de La Serena, 1700000 La Serena, Chile}

\author[0000-0001-8600-4798]{Carlos G. Rom\'an-Z\'u\~niga}
\affil{Instituto de Astronom\'{i}a, Universidad Aut\'{o}noma de M\'{e}xico,
Ensenada, B.C, M\'{e}xico}

\author{Maria Gracia Batista}
\affiliation{Observatorio Astron\'omico Nacional, Facultad de Ciencias, Universidad Nacional de Colombia,
Bogot\'a, Colombia}

\author[0000-0001-9147-3345]{Giovanni Pinz\'on}
\affiliation{Observatorio Astron\'omico Nacional, Facultad de Ciencias, Universidad Nacional de Colombia,
Bogot\'a, Colombia}

\author[0000-0002-3950-5386]{Nuria Calvet}
\affiliation{Department of Astronomy, University of Michigan, 1085 South University Avenue, Ann Arbor, MI 48109, USA}

\author[0000-0001-7124-4094]{Cesar Brice\~{n}o}
\affiliation{Cerro Tololo Inter-American Observatory, Casilla 603, La Serena 1700000, Chile}

\author[0000-0002-0506-9854]{Mauricio Tapia}
\affil{Instituto de Astronom\'{i}a, Universidad Aut\'{o}noma de M\'{e}xico,
Ensenada, B.C, M\'{e}xico}

\author[0000-0002-2011-4924]{Genaro Su\'arez}
\affil{Department of Physics and Astronomy, The University of Western Ontario, 1151 Richmond St, London, ON N6A 3K7, Canada}

\author[0000-0002-5855-401X]{Karla Pe\~na Ram\'irez}
\affil{Centro de Astronom\'ia (CITEVA), Universidad de Antofagasta, Av. Angamos 601, Antofagasta, Chile.}

\author[0000-0002-5365-1267]{Keivan G.\ Stassun}
\affil{Department of Physics and Astronomy, Vanderbilt University, VU Station 1807, Nashville, TN 37235, USA}

\author[0000-0001-6914-7797]{Kevin Covey}
\affil{Department of Physics and Astronomy, Western Washington University, 516 High St, Bellingham, WA 98225}

\author[0000-0003-4329-3299]{J. Vargas-Gonz\'{a}lez}
\affil{Centre for Astrophysics Research, School of Physics, Astronomy and Mathematics, University of Hertfordshire, College Lane, Hatfield AL10 9AB, UK.}

\author[0000-0003-3526-5052]{Jos\'e G. Fern\'andez-Trincado}
\affil{Instituto de Astronom\'ia, Universidad Cat\'olica
del Norte, Av. Angamos 0610, Antofagasta, Chile}
\affil{Instituto de Astronom\'ia y Ciencias Planetarias,
Universidad de Atacama, Copayapu 485, Copiap\'o, Chile}

\begin{abstract}
\noindent We present a large-scale study of stellar rotation for T Tauri stars in the Orion Star-Forming Complex. We use the projected rotational velocity ($v\sin(i)$) estimations reported by the APOGEE-2 collaboration as well as individual masses and ages derived from the position of the stars in the HR diagram, considering Gaia-EDR3 parallaxes and photometry plus diverse evolutionary models. We find an empirical trend for $v\sin(i)$ decreasing with age for low-mass stars ($0.4 M_{\odot}<M_{\ast}<1.2 M_{\odot}$). Our results support the existence of a mechanism linking $v\sin(i)$ to the presence of accreting protoplanetary disks, responsible for regulating stellar rotation in timescales of about 6 Myr, which is the timescale in which most of the T Tauri stars lose their inner disk.
Our results provide important constraints to models of rotation in the early phases of evolution of young stars and their disks.

\end{abstract}


\section{Introduction} 
\label{sec:intro}

\noindent Stellar rotation is a fundamental parameter in star formation and stellar evolution, and it plays a crucial role in the origin of planetary systems. Understanding the origin and early evolution of stellar angular momentum is one of the most challenging problems of modern stellar astrophysics. This study is a cross-sectional tool for multiple stellar physical processes \citep{Bouvier2013, Bouvier2014, Amard2019, Spada_2020}.\\

\noindent While rotational velocities of most main-sequence stars decrease with age ($v\sin(i)\propto t^{-\frac{1}{2}}$) due to angular momentum (AM) losses by stellar winds \citep{Skumanich1972}, in very young stars there are mechanisms not yet well understood that can affect the angular momentum  \citep[e.g., disk-locking effect, core-envelope decoupling, stellar winds;][]{littlefair2013}. Thus, finding Skumanich-type relations for pre-main-sequence stars (PMS) becomes more problematic. Regardless of these complexities, and based on observations from different young stellar associations, it has been possible to progress in AM evolutionary models for low-mass stars at the PMS phase \citep{Matt2012,Gallet2013,Gallet_2015,Amard2016,Amard2019}. The characterization of critical parameters related to the interaction between a star and its protoplanetary disk, mass accretion rate, stellar winds, and other mechanisms of AM loss, needs to be improved to better understand AM evolution during the early stages \citep{Bouvier2014}, when disk dissipation and planet formation occur \citep[e.g.][]{Hernandez2007,Williams2011,Testi2014}.\\

\noindent AM evolutionary models consider a disk-locking mechanism that keeps a constant stellar angular velocity as long as stars have accreting disks \citep{Gallet2013,Gallet_2015,Amard2016,Amard2019}. However, accretion-powered stellar wind models suggest that the stellar angular velocity could change even in the presence of an accreting disk \citep{Matt2012,Gallet2019,Pinzon2021}. Most observational studies support a scenario in which magnetic star-disk coupling has a fundamental impact on young stars rotational properties. Thus, stars with accreting disks rotate, on average, slower than stars without accreting disks \citep[e.g.,][]{Rebull2006a,Jayawardhana_2006,Biazzo_2009,Rebull_2014,Davies2014,Venuti2017}. In contrast, other studies found no significant differences between rotation properties and the presence of accretion disks, even reporting contradictory results to the disk-locking scenario \citep{Stassun1999,Makidon2004,Nguyen_2009,Le_Blanc_2011,Karim2016}. 
Thus, additional studies are needed to determine the role of accreting disks, stellar magnetic fields, and stellar winds on the evolutionary trends of angular momentum observed in PMS stars.\\

\noindent The Orion Star-Forming Complex (OSFC) is one of the best known young regions in the solar neighborhood. It is located at a distance of $d\sim 400$ pc \citep{Grossschedl2018,Kuhn_2019} and has a wide variety of stellar populations in different environments, with ages spanning from 1 to 10 Myr \citep{Briceno2008,Kounkel_2017,Kounkel_2018,Zari2019,Briceno_2019}. The OSFC has several well-known stellar clusters such as the Orion Nebula Cluster (ONC; 1-2 Myr), the $\sigma$ Orionis cluster (3-4 Myr), the $\lambda$ Orionis cluster (4-6 Myr) and the 25 Ori cluster (7-10 Myr), mainly located in two OB associations, the Orion OB1 association \citep{Warren_1977} and the $\lambda$ Orionis association \citep{Murdin_1977}. This makes the OSFC an ideal astrophysical laboratory to perform general studies of star formation and early stellar evolution. 
Thus, the OSFC is well suited to perform a systematic study of stellar rotation during the first million years of the stellar life, when protoplanetary disks with magnetospheric accretion can affect the AM evolution.\\

\noindent Several works have focused on detecting and characterizing young stellar objects (YSOs) on/off clouds of the OSFC. \citet{Briceno_2019} identified more than 2000 T Tauri stars in the Orion OB1 association using optical spectra.
The detection of the \ion{Li}{1} line in absorption confirms the youth of the sample, while measurements of the equivalent width of the H$\alpha$ emission line were used to identify stars with accreting disks (e.g., Classical T Tauri Stars; CTTS), as well as those without accreting disks (e.g., Weak-lined T Tauri Stars; WTTS).
Using similar methods for characterizing young stellar populations,
\citet{Hernandez_2014} and Hernandez et al.(in preparation)
present exhaustive spectroscopic censuses of the stellar population in the the $\sigma$ Orionis cluster and in the ONC, respectively. In addition, combining the second data release of the Gaia space mission \citep[Gaia-DR2;][]{Gaia_2018} and Apache Point Observatory Galactic Evolution Experiment \citep[APOGEE;][]{Majewski2017} high resolution near infrared spectra obtained in the OSFC \citep{Cottaar2014}, \citet{Kounkel_2018} have detected $\sim$2400 kinematic members using 6 dimensional analysis (positions, parallax and proper motions from Gaia-DR2 and radial velocities from APOGEE; hereafter known kinematic members). \\

\noindent Photometric variability is widely used to identify and characterize YSOs \citep[e.g.;][]{Bouvier1993,Herbst_2002,Briceno2008,Morales_Calderon_2011,Cody2018}. In addition to periodic light curves (LCs) modulated by stellar rotation and the presence of magnetic spots on a stellar surface, other mechanisms contribute to the observed variability in YSOs such as dimming by the dust in the disk, fluctuations in the accretion rate of the disk, eclipses caused by stellar companions or planets, and stellar pulsations \citep{Cody2010,Cody2014}. Nowadays, space missions have revolutionized the sky photometric monitoring, providing unprecedented information on the variable sky. Particularly, the Transiting Exoplanet Survey Satellite (TESS), which has finished its second year of operations providing very high-quality photometric data, is one of the most powerful large-scale astrophysical space missions. It focuses primarily on detecting transits of planets orbiting bright host stars relatively close to the earth \citep{Ricker2014}. TESS has provided accurate photometry with a cadence of 30 minutes during 27 days of observations for the entire OSFC. In this work, we use TESS data to build LCs and measure rotational periods of T Tauri stars, which normally range from 1 to 10 days \citep{Karim2016}.\\

\noindent Making use of large-scale spectroscopic, photometric, and kinematic surveys in the OSFC \citep{Kounkel_2018,Briceno_2019}, we perform here an evolutionary study of the stellar rotation for low-mass stellar members ($0.4 M_{\odot}<M_{\ast}<1.2 M_{\odot}$) in this star-forming complex. We investigate stellar rotation in T Tauri stars and the relation to their protoplanetary disks. We also provide new rotational periods for kinematical and spectroscopic members of the OSFC, derived from TESS data. In \S \ref{sec:data}, we define our sample of spectroscopic and kinematic members. We provide a brief description of the TESS observations and the LC analysis extraction, and we also describe the derivation and compilation of the stellar parameters such as $v\sin(i)$, effective temperature ($T_{eff}$), rotation period ($P_{\rm rot}$), mass, and age. In \S \ref{sec:results} we present the derived relationships of $v\sin(i)$ against period, $v\sin(i)$ versus age, and $v\sin(i)$ versus H$\alpha$ line equivalent width ($EW_{H_{\alpha}}$). Also, we discuss the influence of disks on the rotation measurements of the OSFC, including evidence of disk-locking effects. Finally, in \S \ref{sec:summary} we summarize the results and present our main conclusions.\\

\section{OSFC sample and measurements of stellar parameters}
\label{sec:data}

\subsection{Kinematic Members}
\label{sec:members2}

\noindent The Gaia mission has opened the possibility of detecting young stars by studying their kinematic properties with unprecedented precision and accuracy \citep{Kounkel_2018,Zari2019,Kuhn_2019,Soubiran_2019,GodoyRivera2021}. Recently, the Early Third Data Release (hereafter Gaia-EDR3) of the Gaia mission became available. It includes significant improvements with respect to previous releases \citep{GAIA_EDR3}.
We cross-matched the known kinematic members reported by \citet{Kounkel_2018} with the Gaia-EDR3 catalog.  Based on the distributions of Gaia-EDR3 parallaxes ($\varpi$) and proper motion modulus ($|\mu|=\sqrt{\mu_\alpha^2+\mu_\delta^2}$) of the known kinematic members, we define the following membership region: 2.00$<\varpi<$ 3.14 mas and $|\mu|<$3.5 mas/yr. These limits were defined  using the mean and the standard deviation applying a 3$\sigma$ criteria. Our sample of kinematic members includes stars with APOGEE infrared spectra \citep{Kounkel_2018,Cottle2018,Majewski2017}  that fall in the membership region. Thus, we include stars with reliable astrometric solutions in Gaia-EDR3, not included in the previous Gaia release.
We require stars with relative errors in parallax smaller than 20\%. We also apply cuts based on the photometric and astrometric quality suggested for Gaia-EDR3 data \citep{Lindegreen2020a}. In brief, we select all kinematic members with renormalized unit weight error (RUWE) smaller than 1.4 and satisfying the \citet{Lindegreen2020a} relation:
\begin{equation}
\label{Eratio}
0.001 + 0.039\times(BPmag - RPmag) < \log E < 0.12 + 0.039\times(BPmag - RPmag)
\end{equation}

\noindent where E is the excess flux ratio \footnote{$E=\frac{I_{BP}+I_{RP}}{I_G}$ ; where $I_{BP}$,$I_{RP}$ and $I_G$ are the fluxes in the Gaia filters \textit{BP}, \textit{RP} and \textit{G}, respectively} and \textit{BPmag} and \textit{RPmag} are the magnitudes in the filters \textit{BP} and \textit{RP}, respectively. Kinematic members that do not fulfill equation (\ref{Eratio}) could have inconsistent Gaia-EDR3 photometric fluxes due to blends, contamination by a nearby source or a sign of the extended nature of the source. Finally, using the {\sc gaiadr3\_zeropoint} python package,  we estimate the value of the parallax zero-point for Gaia-EDR3. This software applies the analytical functions to compute the expected parallax zero-point as a function of ecliptic latitude, magnitude and color \citep{Lindegreen2020b}. To reduce systematics, we subtract the parallax zero-point from the Gaia-EDR3 parallaxes.
The left panel of Figure \ref{fig:orion} shows the spatial distribution of the kinematically selected YSOs in the complex. 
The resulting sample has 2204 kinematic members of the Orion Complex that will be used in this work.

\subsection{Spectroscopic Members}
\label{sec:members1}

\begin{figure}[htp]
    \centering
    \includegraphics[width=0.33\textwidth]{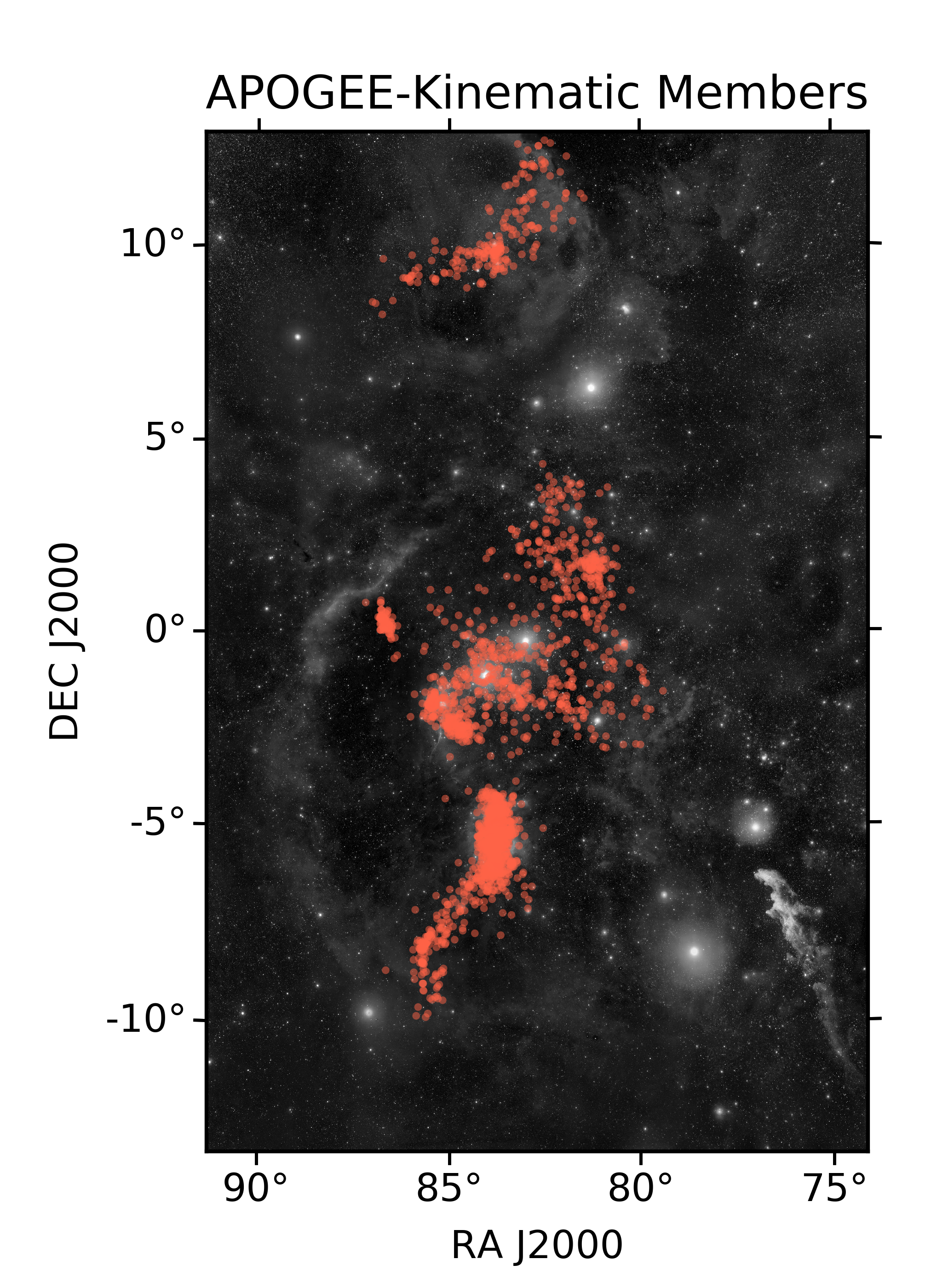}\hfill
    \includegraphics[width=0.33\textwidth]{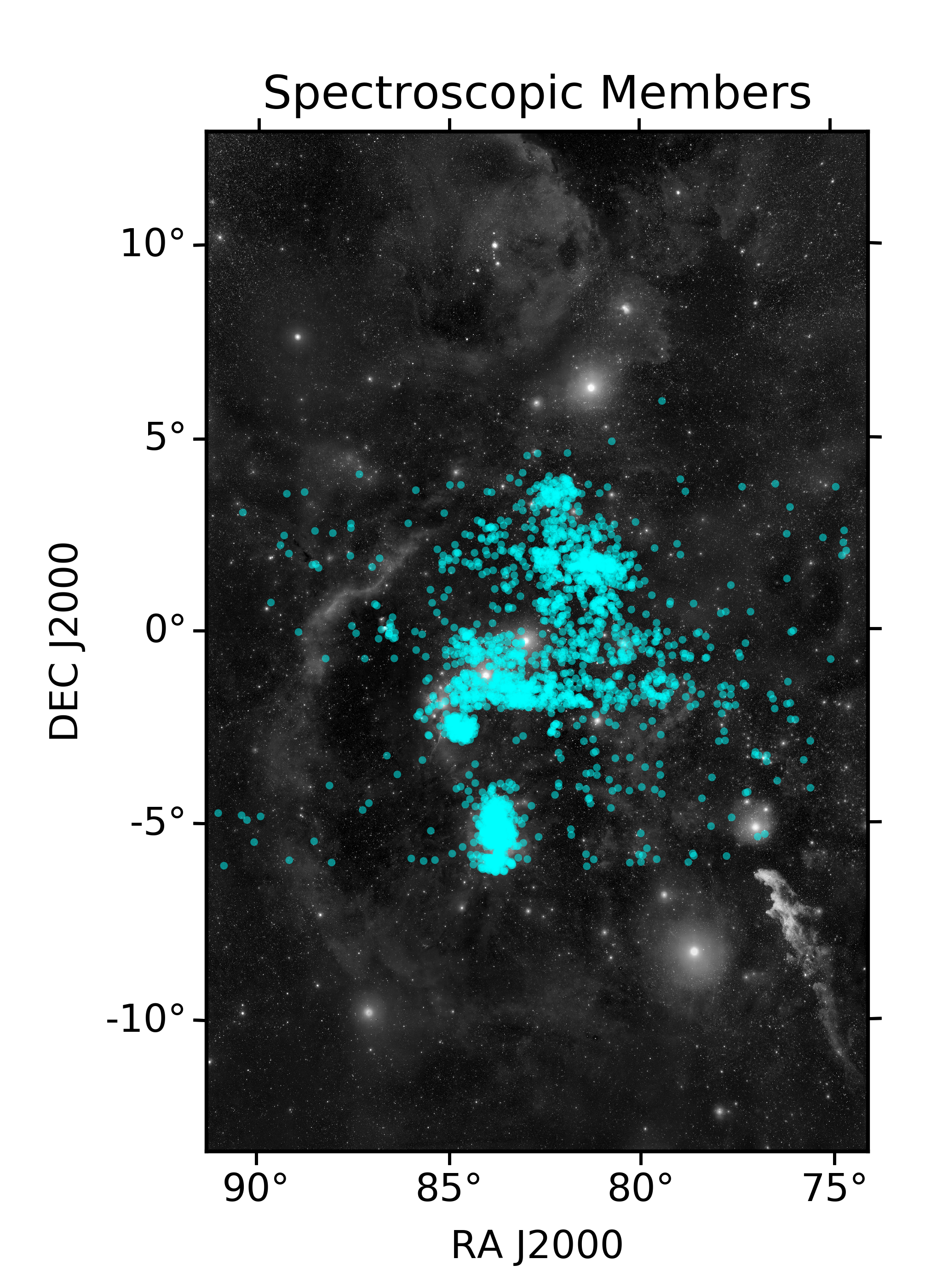}\hfill
    \includegraphics[width=0.33\textwidth]{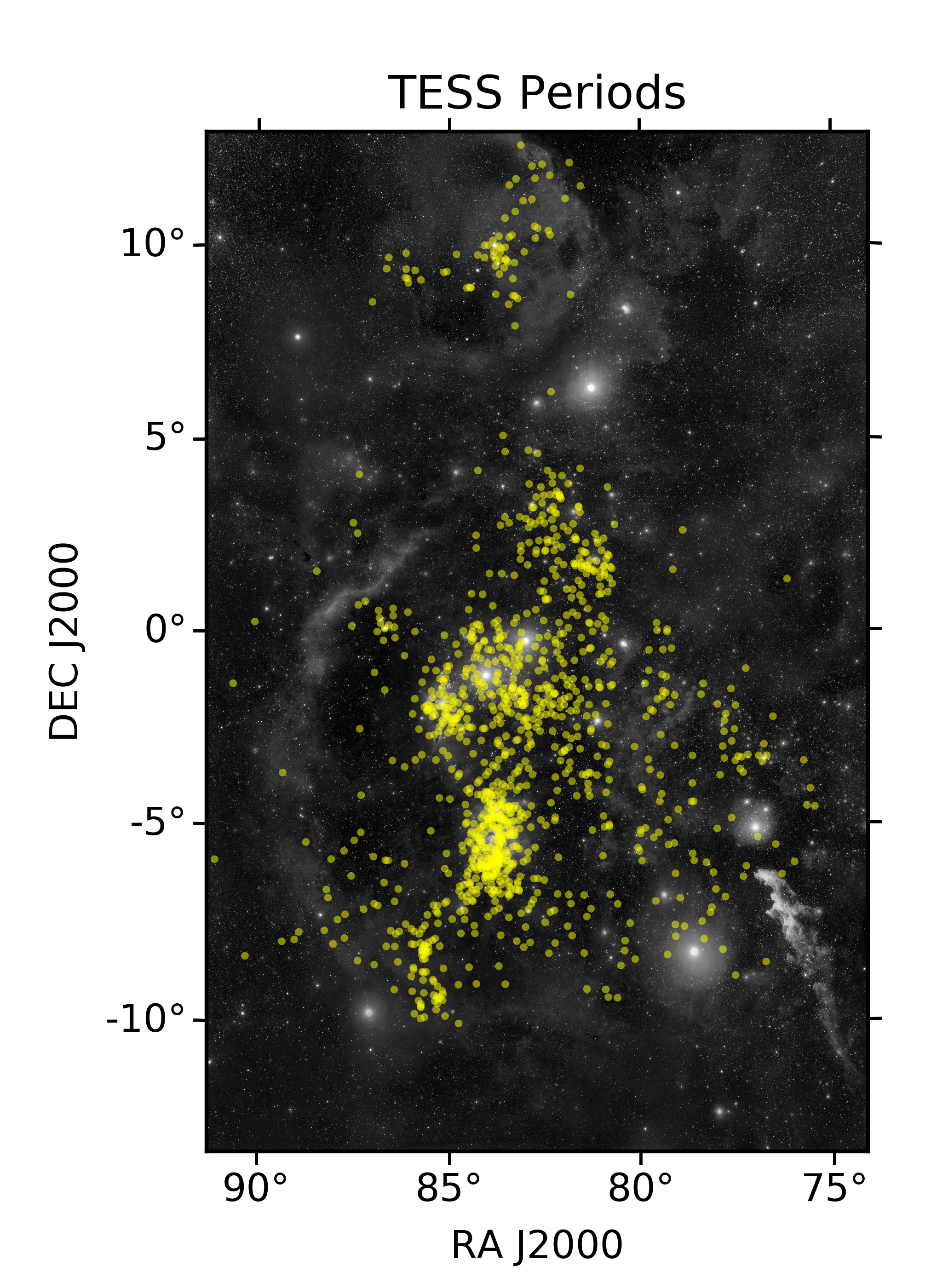}
    \caption{Spatial distribution of Orion's APOGEE sources with counterparts in the kinematic members sample (red dots in the left panel), spectroscopic members (middle panel, cyan dots), and kinematic or spectroscopic members with TESS LCs (right panel, yellow dots) in the OSFC sample. The gaps observed in the middle panel are due to a selection effect of the spectroscopic sample \citep[see, ][]{Briceno_2019}. The background image is an astrophotograph in optical filters as reference, courtesy of Rogelio Bernal Andreo (modified to grayscale).}
    \label{fig:orion}
\end{figure}

\noindent An extensive optical spectroscopic study in
the Orion OB1 associations was performed by \citet{Briceno_2019}. Using several instruments with similar spectral resolution (R$\sim$ 1000-2000), they obtained optical spectra for a sample of 11200 candidate PMS stars selected by their photometric variability and location in optical and optical-near-IR color-magnitude diagrams. Using the IRAF/IDL based SPTCLASS tool \citep{Hernandez2004,Hernandez2017}, \citet{Briceno_2019} analyzed the mentioned sample of candidates and reported 2062 T Tauri Stars (TTSs) that were listed with their spectral types and equivalent widths of \ion{Li}{1} $\lambda$6708{\AA} and H$\alpha$ lines. 
Similarly, and using the same tools, \citet{Hernandez_2014} and Hernandez et al., (in preparation) performed a systematic optical spectroscopic census of YSOs, reporting 221 TTS in the $\sigma$ Orionis cluster and 909 TTS in the ONC \citep[see also][]{Manzo_2020}. The middle panel of Figure \ref{fig:orion} shows the spatial distribution of the spectroscopically-confirmed TTS.

\noindent Using a classification scheme based on the relation between the equivalent widths of the H$\alpha$ line and spectral types, \citet{Briceno_2019} reported 1696 WTTS, 214 CTTS, and 152 CWTTS in the Orion OB1 association.
The CWTTS are considered stars evolving from an active CTTS accretion phase to a non-accreting WTTS stage. \citet{Briceno_2019} proposed that CWTTS is likely composed of a mix of objects that are accreting at modest or low levels, constituting the weak tail of the CTTS and a few objects in a quiescent stage between periods of enhanced accretion.
In this work, we apply the same classification scheme to split the $\sigma$ Ori and the ONC samples of TTS into WTTS, CTTS, and CWTTS. This information about the accretion status of spectroscopically confirmed TTS is crucial to examine the relation between the stellar rotational properties and the presence of accreting protoplanetary disks (see \S \ref{sec:rothalpha}).

\subsection{Rotational Velocity and Effective Temperature}
\label{sec:vsini}

\noindent The effective temperatures ($T_{eff}$), surface gravities ($\log g$), and radial velocities (RV) of the stars in the OSFC sample \citep{Cottle2018} were reported by \citet{Kounkel_2018} using the IN-SYNC pipeline \citep{Cottaar2014}\footnote{The $v\sin(i)$ for this sample was reported in \citet{Kounkel_2019}}. IN-SYNC was designed to analyze APOGEE spectra of YSOs as an alternative to the APOGEE Stellar Parameter and Chemical Abundances Pipeline \citep[ASPCAP;][]{Garcia-Perez_2016}, which is primarily optimized for giants stars. The IN-SYNC code fits to each APOGEE spectrum $T_{eff}$, $\log g$, RV, veiling and $v\sin(i)$, using a synthetic grid, and interpolating between the grid points \citep{Kounkel_2018}.\\ 
\noindent \citet{Olney_2020} find that the stellar parameters estimated by \citet{Kounkel_2018} could still include some non-physical systematics due to mismatches between the empirical and theoretical spectra. Thus, using a deep convolutional neural network, the APOGEE-net pipeline was designed to correct most of these issues, providing more reliable predictions of $\log g$, $T_{eff}$ and [Fe/H] for solar and lower mass regime \citep[$T_{eff}\lesssim6700$ K;][]{Olney_2020}.

\noindent Since \citet{Olney_2020} studied only stars within the APOGEE DR14 data set, not all APOGEE fields in the OSFC and the ones studied in \citet{Kounkel_2018, Kounkel_2019} were included in their work. 
Thus, we use the APOGEE-net to obtain the best estimates for the stellar parameters of the entire kinematic members selected in \S \ref{sec:members2}. Since the APOGEE-net does not predict $v\sin(i)$, we made use of the $v\sin(i)$ values obtained from \citet{Kounkel_2019}. In general, the differences between the stellar parameters determined in \citet{Kounkel_2018} with the IN-SYNC pipeline and the APOGEE-net pipeline are less than 500 K for $T_{eff}$ and less than 1.0 dex for $\log g$.
To investigate whether these differences could produce systematic errors in the $v\sin(i)$, we made a quality check for $v\sin(i)$ values reported in \citet{Kounkel_2019}, using the Fourier Transform (FT) method, which does not require the use of spectroscopic templates \citep{Carroll1933}.\\

\noindent The FT method is widely used to study stellar rotation in different ranges of stellar masses \citep{Royer_2002,Simon-Diaz2007,Thanathibodee_2020}. In short, the observed star's spectrum can be written as a convolution of the intrinsic stellar spectrum, the broadening function, and the instrumental function \citep{Carroll1933,Royer2005}. The rotational broadening function produces zeros in the FT space in which the first zero is correlated with $v\sin(i)$. Other broadening functions as Stark, thermal, macro-turbulence broadening do not create zeroes in the FT space, and do not affect the derivation of $v\sin(i)$ \citep{Carroll1933,Royer2005}. The instrumental function does not add zeroes in the FT space but it is related to the lower limit of $v\sin(i)$ that we can measure in a given spectral resolution.

\noindent In order to estimate the rotational velocity using the FT method, an interactive tool was designed in Python and Qt as a part of a suite of tools for stellar rotation (Serna et al., in preparation).
To obtain robust estimations of $v\sin(i)$, the FT method requires the selection of single spectroscopic lines (not blended), with good signal to noise (S/N) to get a reliable normalization \citep{Royer2005}. The tool enables the user to select and analyze the shape of the spectral line of interest. The tool automatically gets the line center and width by fitting a Gaussian function to the spectral line and the continuum using the routine \textit{lmfit} \citep{Newville_2014}.
Thus, the line profile is properly normalized and convolved using the Fast Fourier Transform (FFT) algorithm to get zeroes at the FT space. We use only the first zero ($\sigma_{1}$), which, with the APOGEE spectral resolution, provides reliable $v\sin(i)$ \citep{Reiners2002}. Thus, the rotational velocity is computed as $v\sin(i) = \frac{k_{1}(\epsilon)}{\sigma_{1}}$. Following \citet{Reiners2002} prescription, we use the following approximation based on the limb-darkening coefficient ($\epsilon$):
$k_{1}(\epsilon)= 0.610 + 0.062\epsilon + 0.027\epsilon^{2} + 0.012\epsilon^{3} + 0.004\epsilon^{4}$.
To evaluate the robustness of the method and to obtain uncertainties for our measurements, the tool creates an ensemble of synthetic lines for each measure, resampling 1000 times the input line, taking into account the uncertainties of the flux and the wavelength, with the limb-darkening coefficient varying randomly from 0.4 to 0.6 \citep{Gilhool2018}. The program then computes $v\sin(i)$ from the median value and the associated uncertainty from the standard deviation. \\

\begin{figure}[htp]
    \centering
    \includegraphics[width=0.45\textwidth]{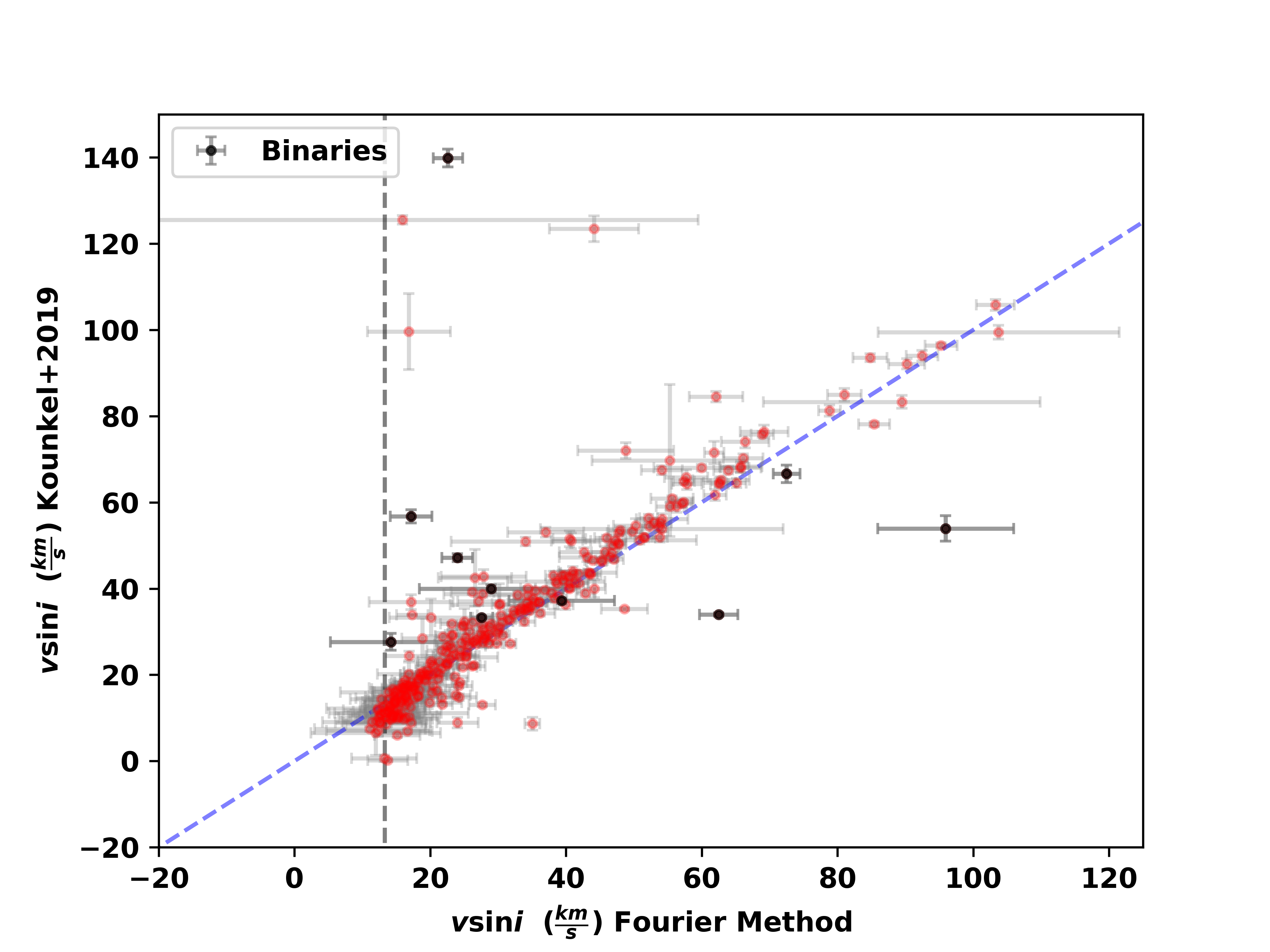}
    \includegraphics[width=0.45\textwidth]{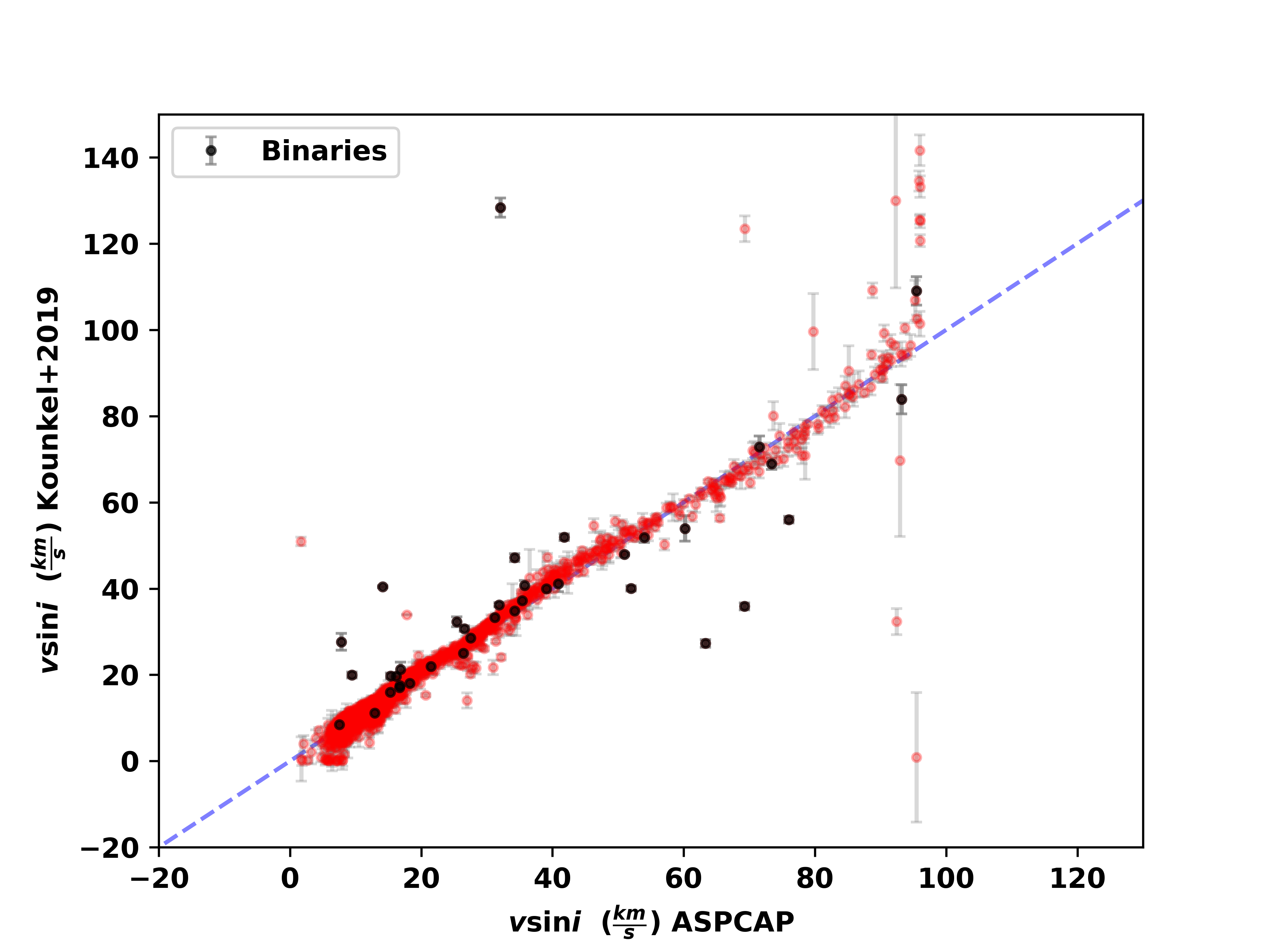}
    \includegraphics[width=0.45\textwidth]{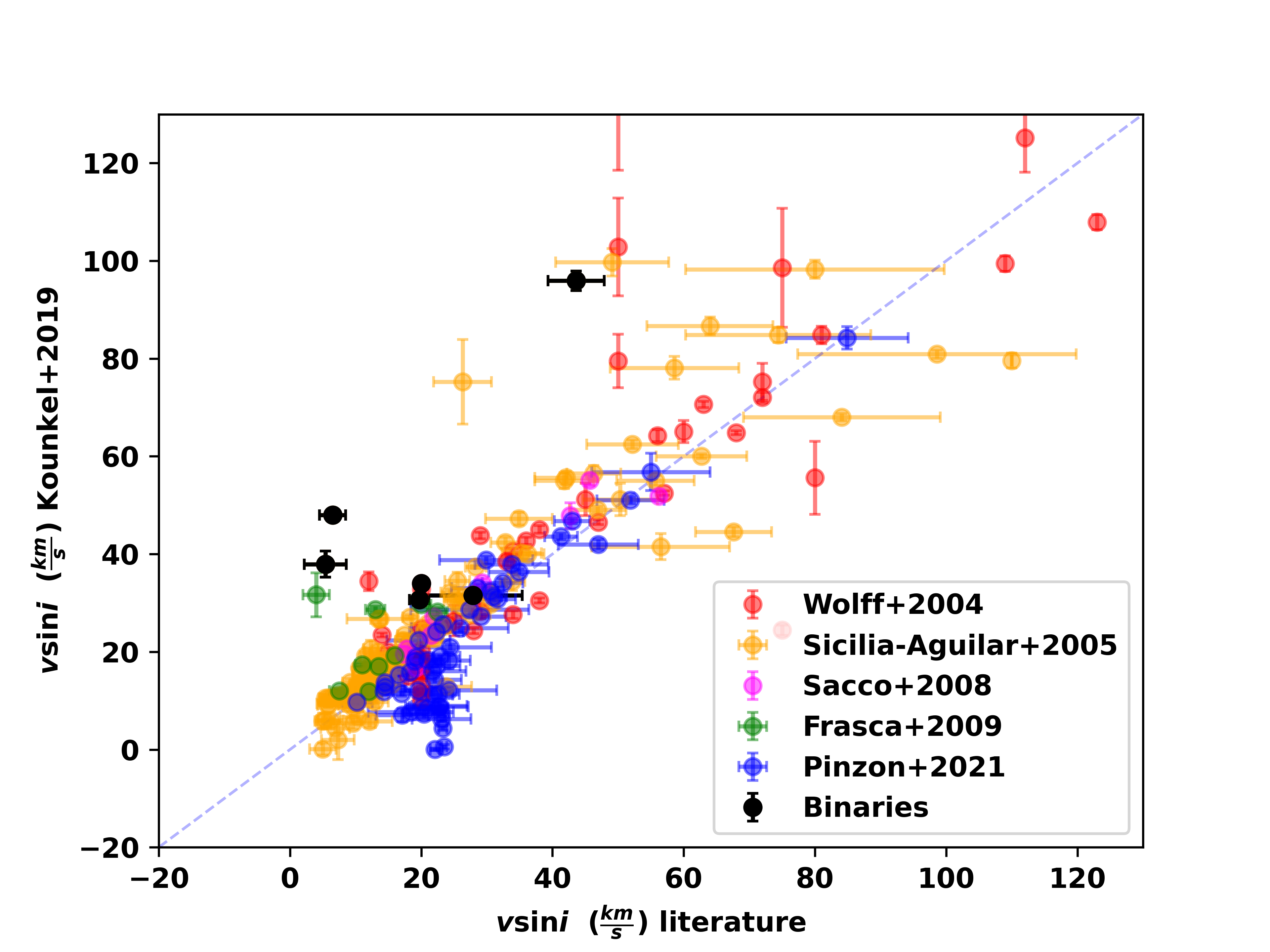}
    \caption{Quality check for v$\sin(i)$ measurements:
    The upper left panel shows a validation between the $v\sin(i)$ measurements of \citet{Kounkel_2019},
    which we used in this work, and the measurements that we made with the Fourier Method. The upper right panel shows a comparison of \citet{Kounkel_2019} versus ASPCAP. Notice that the ASPCAP does not estimate $v\sin(i)$ larger than 100 km~s$^{-1}$. The lower panel shows a comparison of \citet{Kounkel_2019} versus other works in the literature. Black dots denote confirmed spectroscopic binaries \citep{Kounkel_2019}. The slope of the blue dashed line in each box is one, and the vertical black dashed line in the upper left box marks the APOGEE velocity resolution $\sim13.3$ km~s$^{-1}$.}
    \label{fig:comparison}
\end{figure}

\noindent To guarantee proper line shapes in the FT analysis,
we have measured $v\sin(i)$ for 271 kinematic members (\S \ref{sec:members2}) with APOGEE spectra presenting S/N$>$200. 
We analyzed at least 5 of the most intense lines within the APOGEE spectroscopic range for each star, avoiding lines with apparent mixing with other lines. The final values (the median of the set of individual results) are shown in Table \ref{tab:siniperiodos} \footnote{Appendix \ref{apendiceA} includes velocity measurements for stars located in the general region of the OSFC not included as kinematic (\S\ref{sec:members2}) or spectroscopic(\S\ref{sec:members1}) members}. 
All measurements with $v\sin(i)<13.3$ km~s$^{-1}$ provided by FT method are considered upper limits. The left panel of Figure \ref{fig:comparison} compares the $v\sin(i)$ derived by \citep{Kounkel_2019} and the $v\sin(i)$ estimated from the FT method. In general, both sets of measurements, above the APOGEE velocity resolution, are in good agreement within 5 km~s$^{-1}$. Similarly, the right panel of Figure \ref{fig:comparison} compares the $v\sin(i)$ derived by \citep{Kounkel_2019}, to those derived using the ASPCAP pipeline, which uses different spectral theoretical libraries and analyzes the combined spectra obtained from multi-epoch APOGEE observations \citep{Garcia-Perez_2016}. The comparison shows a good agreement and does not reflect any biases; We also made some comparisons with the literature \citep{Wolff_2004,Sicilia_Aguilar_2005,Sacco_2008,Frasca_2009,Pinzon2021}. In general, we have a good agreement with \citet{Kounkel_2019}.

\noindent Most stars with the larger differences in Figure \ref{fig:comparison} are spectroscopic binaries detected by \citet{Kounkel_2019}. The $v\sin(i)$ estimation in binary stars can be affected by blended spectroscopic features from the binary components. 
Since we can not obtain reliable $v\sin(i)$ in stars with low signal to noise (S/N$<$200) using the FT method and ASPCAP $v\sin(i)$ are available only for stars in the APOGEE DR14 data set, we use in the following sections the $v\sin(i)$ values derived by \citet{Kounkel_2018} and reported in \citet{Kounkel_2019}.\\

\begin{longrotatetable}
\begin{deluxetable*}{cccccccccccc}
\tablecaption{OSFC rotational velocities and rotational periods. \label{tab:siniperiodos}}
\tablehead{
\colhead{2massID}   &  \colhead{$v\sin(i)$\tablenotemark{a}} & \colhead{$v\sin(i)$\tablenotemark{b}}  & \colhead{Binary\tablenotemark{a}} & \colhead{Kinematic} &\colhead{Accretor} & \colhead{EW[H$_\alpha$]} &\colhead{TIC} & \colhead{Tmag} &\colhead{Period\tablenotemark{c}} & \colhead{Period\tablenotemark{d}} &\colhead{References\tablenotemark{d}} \\
\colhead{       }   &  \colhead{km/s}   & \colhead{km/s}    & \colhead{      } & \colhead{         } &\colhead{Class} & \colhead{\AA}   &\colhead{ } & \colhead{mag} &\colhead{days} & \colhead{days} &\colhead{}
}
\startdata
   05180280-0106222  &    16.7 $\pm$   0.7  &  \nodata  &	1	&  \nodata  &   W  &     -5.3  &  \nodata  &  \nodata  &  \nodata  &  \nodata  &  \nodata \\
  05181171-0001356  &  \nodata  &  \nodata  &	\nodata	&  \nodata  &   W  &     -3.7  &   249067805  &   13.16  &    1.65 $\pm$  0.01 	 &  1.65  &  7 \\
  05181457+0010095  &  \nodata  &  \nodata  &	\nodata	&  \nodata  &   W  &     -5.6  &   454223248  &   14.21  &    6.70 $\pm$  0.03 	 &  \nodata  &  \nodata \\
  05185716-0256047  &    78.3 $\pm$   6.0  &  \nodata  &	0	&   1  &  \nodata  &  \nodata  &  \nodata  &  \nodata  &  \nodata  &  \nodata  &  \nodata \\
  05190273-0032274  &  \nodata  &  \nodata  &	\nodata	&  \nodata  &   W  &     -7.7  &   249074941  &   14.79  &    8.23 $\pm$  0.01 	 &  8.21  &  7 \\
  05191214-0242275  &    13.5 $\pm$   0.6  &  \nodata  &	1	&  \nodata  &   W  &     -3.6  &  \nodata  &  \nodata  &  \nodata  &  \nodata  &  \nodata \\
  05191549-0204529  &     7.4 $\pm$   1.2  &  \nodata  &	1	&   1  &  \nodata  &  \nodata  &  \nodata  &  \nodata  &  \nodata  &  \nodata  &  \nodata \\
  05192175-0217193  &    16.4 $\pm$   0.6  &  \nodata  &	1	&   1  &  \nodata  &  \nodata  &     4011038  &   13.03  &    4.88 $\pm$  0.01 	 &  \nodata  &  \nodata \\
  05194010-0121224  &    28.2 $\pm$   5.9  &  \nodata  &	0	&   1  &  \nodata  &  \nodata  &  \nodata  &  \nodata  &  \nodata  &  \nodata  &  \nodata \\
  05194349-0116397  &    10.2 $\pm$   0.6  &  \nodata  &	1	&   1  &  \nodata  &  \nodata  &  \nodata  &  \nodata  &  \nodata  &  \nodata  &  \nodata \\
  05194988-0408068  &  \nodata  &  \nodata  &	\nodata	&  \nodata  &   W  &     -5.9  &     4068446  &   14.06  &    0.52 $\pm$  0.01 	 &  \nodata  &  \nodata \\
  05195655-0520227  &  \nodata  &  \nodata  &	\nodata	&  \nodata  &   W  &     -4.2  &     4069106  &   13.75  &    3.36 $\pm$  0.01 	 &  \nodata  &  \nodata \\
  05195766-0301262  &    16.4 $\pm$   0.4  &  \nodata  &	1	&   1  &  \nodata  &  \nodata  &  \nodata  &  \nodata  &  \nodata  &  \nodata  &  \nodata \\
  05200104-0100101  &   198.9 $\pm$  21.8  &  \nodata  &	0	&   1  &  \nodata  &  \nodata  &  \nodata  &  \nodata  &  \nodata  &  \nodata  &  \nodata \\
  05200250-0229098  &     0.0 $\pm$   0.8  &  \nodata  &	1	&   1  &  \nodata  &  \nodata  &  \nodata  &  \nodata  &  \nodata  &  \nodata  &  \nodata \\
\enddata 
\tablenotetext{a}{\citet{Kounkel_2018,Kounkel_2019}: 0- Undeconvolvable cross-correlation function (CCF); 1- Only a single component in the CCF; 2- Multiple components in the CCF; -1- Spotted pairs or SB2 Uncertain}
\tablenotetext{b}{Fourier Method, \S\ref{sec:vsini}}
\tablenotetext{c}{TESS periods \S\ref{sec:tessperiods}}
\tablenotetext{d}{Known periods:1) \citet{Stassun1999}; 2)\citet{Rebull2001,Rebull2006a}; 3) \citet[][;J-band]{Carpenter2001}; 4) \citet{Herbst_2002}; 5) \citet{Cody2010}; 6) \citet{Morales_Calderon_2011}; 7) \citet[7]{Karim2016}}
\tablecomments{Only a portion of the table is shown here. The full version is available in electronic form.}
\end{deluxetable*}
\end{longrotatetable}

\subsection{The TESS Data and Light Curves}
\label{sec:tesslc}

\noindent TESS is an all-sky survey mission focused on searching for transit exoplanets. To date, TESS has covered $\sim75\%$ of the sky with at least 27 days of continuous observation with a cadence of 30 minutes and a pixel size projected into the sky of $21\arcsec\times21\arcsec$. Since TESS was optimized to survey stars in the spectral range F5-M5 \citep{Ricker2014}, these data present a golden opportunity for photometric variability studies in solar-type and low-mass stars.
Particularly, in this work, we use the high quality and cadence of TESS photometric data to estimate stellar rotational periods of TTS in the OSFC, which is widely covered by TESS Sectors 5 \& 6.\\

\noindent We developed the \textit{TESSExtractor} tool specially optimized for periodical variability searches of YSOs using TESS data. It can use as input the Full Frame Images (FFIs) directly, or use the \textit{TESScut} \citep{TESSCut} service to downloads $10\times10$ squared pixels cutouts, centered on each target.\\
To extract and process LCs of selected stars, it selects an optimal aperture to perform simple aperture photometry (SAP) using Python \textit{Photutils} package \citep{Bradley_2019}.
The optimal photometric aperture depends on the star's brightness and can vary from 1.0 to 3.5 pixels radii. The annulus used for the background estimation depends on the selected aperture, with the inner and outer radius varying from 2.5 to 4.0 pixels, and 3.5 to 5.0 pixels, respectively.
Also, the TESS quality flags were used to avoid any anomaly in the photometry (e.g., Cosmic rays, Popcorn noise, Fireworks, etc.)\footnote{TESS Data release notes: \url{https://archive.stsci.edu/tess/tess_drn.html}}.\\ 

\noindent We implemented the task \textit{kepcotrend} of PyKE package \citep{PyKE} in our tool, and the first four Cotrending Basis Vectors (CBVs) provided by TESS to identify and reject LCs dominated by systematics. To avoid possible contamination from scattered light patterns on the TESS detector, we reject all data points with a strong sky variability, above 95$\%$ of the median background estimation.
Finally, each LC was calibrated to the TESS photometric system based on \citet[TIC v8.0]{Stassun2019}.\\

\begin{figure}[htp]
    \centering
    \includegraphics[width=0.7\textwidth]{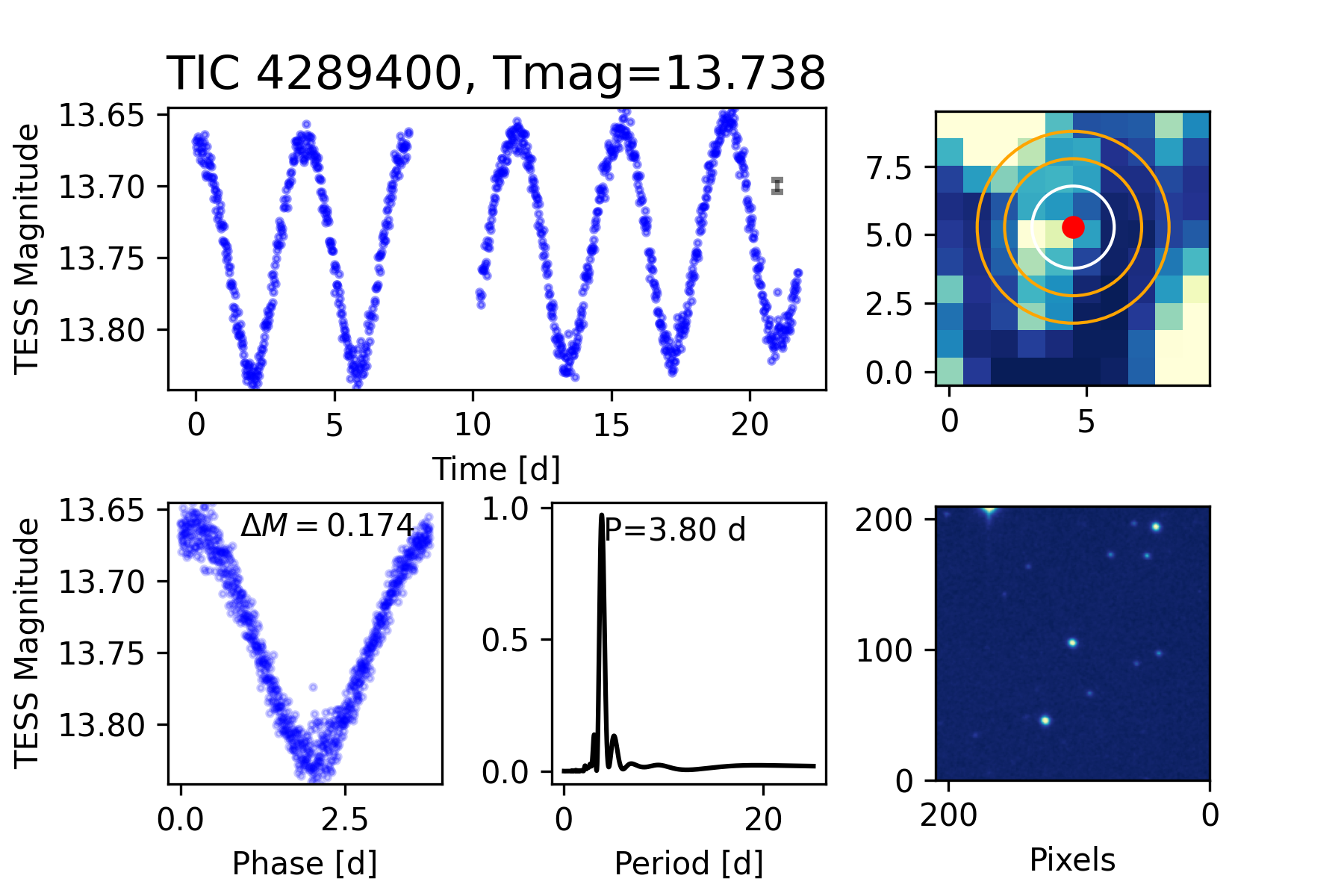}
    \caption{An example of a TESS light curve and its analysis. 
    From the left to the right in top: (A) Light curve in TESS magnitudes with a representative error bar. The legend at top refers to the star identification, and the mean TESS magnitude. (B) Field of view (210 x 210 sq arcsec) corresponding to a TESS image of 10x10 pixels. The white circle shows the photometric aperture, and the orange circles the sky annulus. The red dot marks the centroid of the star. In the bottom panels from the left to the right: (C) Phase-folded light curve to the estimated best period. The legend shows the amplitude. (D) Lomb-Scargle periodogram, with the estimated period. (E) 210 x 210 pixels Digital Sky Survey (DSS2) thumbnail, same field of view as (B).}
    \label{fig:pipe}
\end{figure}

\noindent Given the relatively large TESS pixel size, some LCs may be contaminated by point sources located inside the selected photometric aperture. To consider this issue we computed, the ratio between the flux of the target and the 
the sum of the fluxes of Gaia-sources that fall within the aperture ($\frac{f_{in}}{f_{\ast}}$). The fluxes are based on the G filter of Gaia-EDR3.
 We flag the targets with $\frac{f_{in}}{f_{\ast}}>20\%$
as stars contaminated by nearby sources.
Additionally, to facilitate visual inspection and verify if neighboring sources contaminate the target source, \textit{TESSExtractor} generates an image per each studied target, where one can compare the DSS2 image and TESS image with a field of view of $210\arcsec\times210\arcsec$ corresponding to a TESS image of $10\times10$ pixels (see Figure \ref{fig:pipe}).

\subsubsection{TESS Rotational Periods}
\label{sec:tessperiods}

\noindent 
Using the Lomb-Scargle periodogram \citep[LSP;][]{Lomb_1976,Scargle_1982}  within the \textit{TESSExtractor} tool, we obtain stellar rotational periods for a sample of kinematic members and spectroscopic members in the OSFC (\S \ref{sec:members2} \& \ref{sec:members1}). We use a grid of 1000 periods in the interval of $0.04<P<25$ days. Our minimum period is twice the TESS cadence, and the maximum period of $25$ days is the average length of the TESS time series.
We use the bootstrap technique as a proper way to estimate the period uncertainty. Each LC was re\textit{-}sampled 100 times within the magnitude uncertainties and the rotational period was recomputed in each iteration. At the end, we have an ensemble of measurements, where the reported period is the median of the individual values, and the standard deviation is adopted as the period uncertainty. 
Given the highest peak in the periodogram and assuming Gaussian noise, we have obtained the maximum False Alarm Probability (FAP) of the estimated period, based on the Baluev method \citep{Baluev_2008} using the task \textit{statistics.false\_alarm\_probability} from astropy package \citep{2013A&A...558A..33A}.
As a result, we have FAPs below 0.01\%, indicating that the period measurements are highly reliable.
We perform phase folding of the LCs at the estimated period (Figure \ref{fig:pipe}). Additionally, we select those stars that show a clear periodicity by visual inspection since their LCs show a regular pattern.
Rotational periods are reported for this sample in Table \ref{tab:siniperiodos} \footnote{Appendix \ref{apendiceA} includes TESS rotational periods for stars located in the general region of the OSFC not included as kinematic (\S\ref{sec:members2}) or spectroscopic(\S\ref{sec:members1}) members}. 

\begin{figure}[htp]
    \centering
    \includegraphics[width=0.5\textwidth]{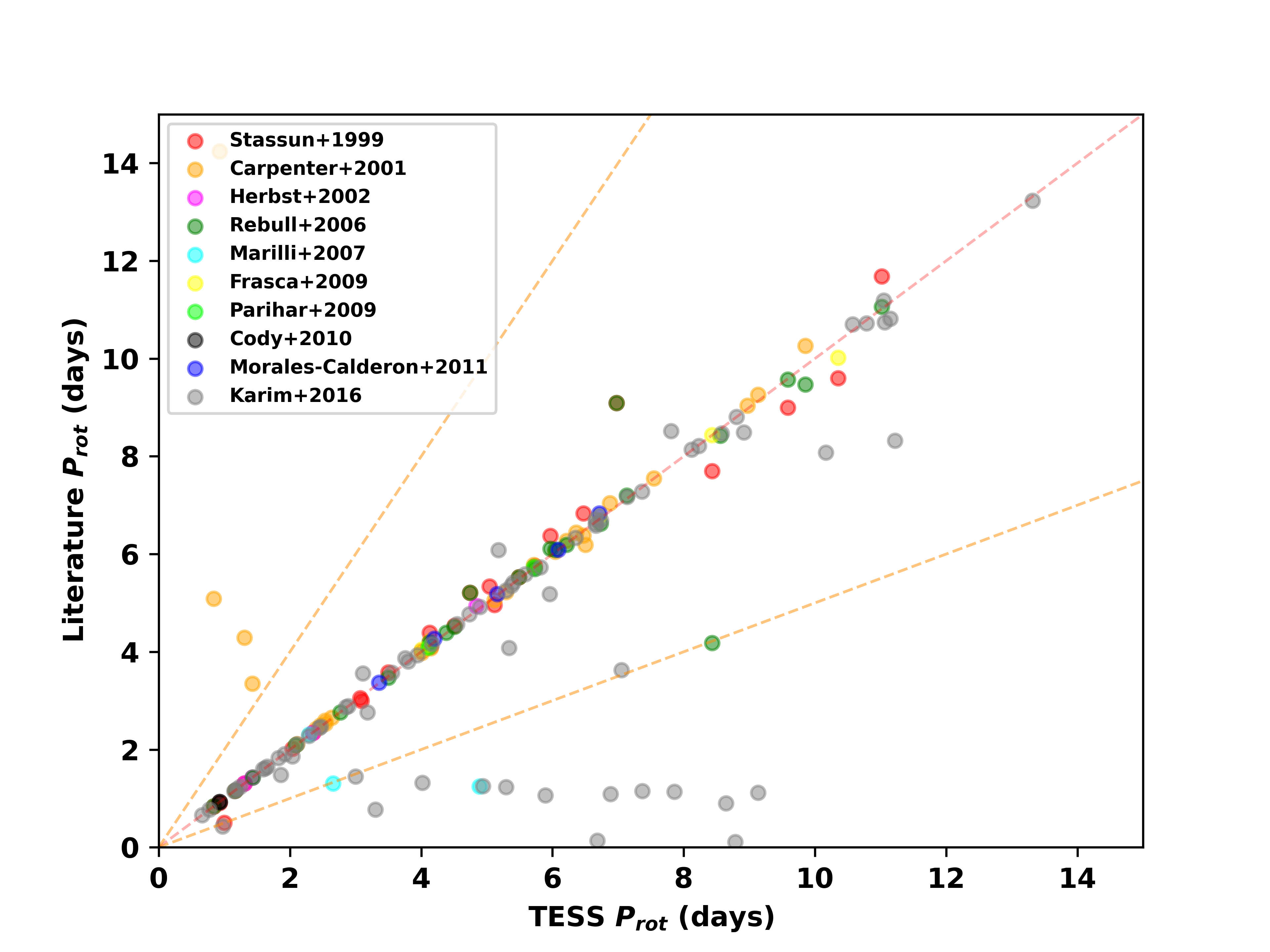}\hfill
    \includegraphics[width=0.5\textwidth]{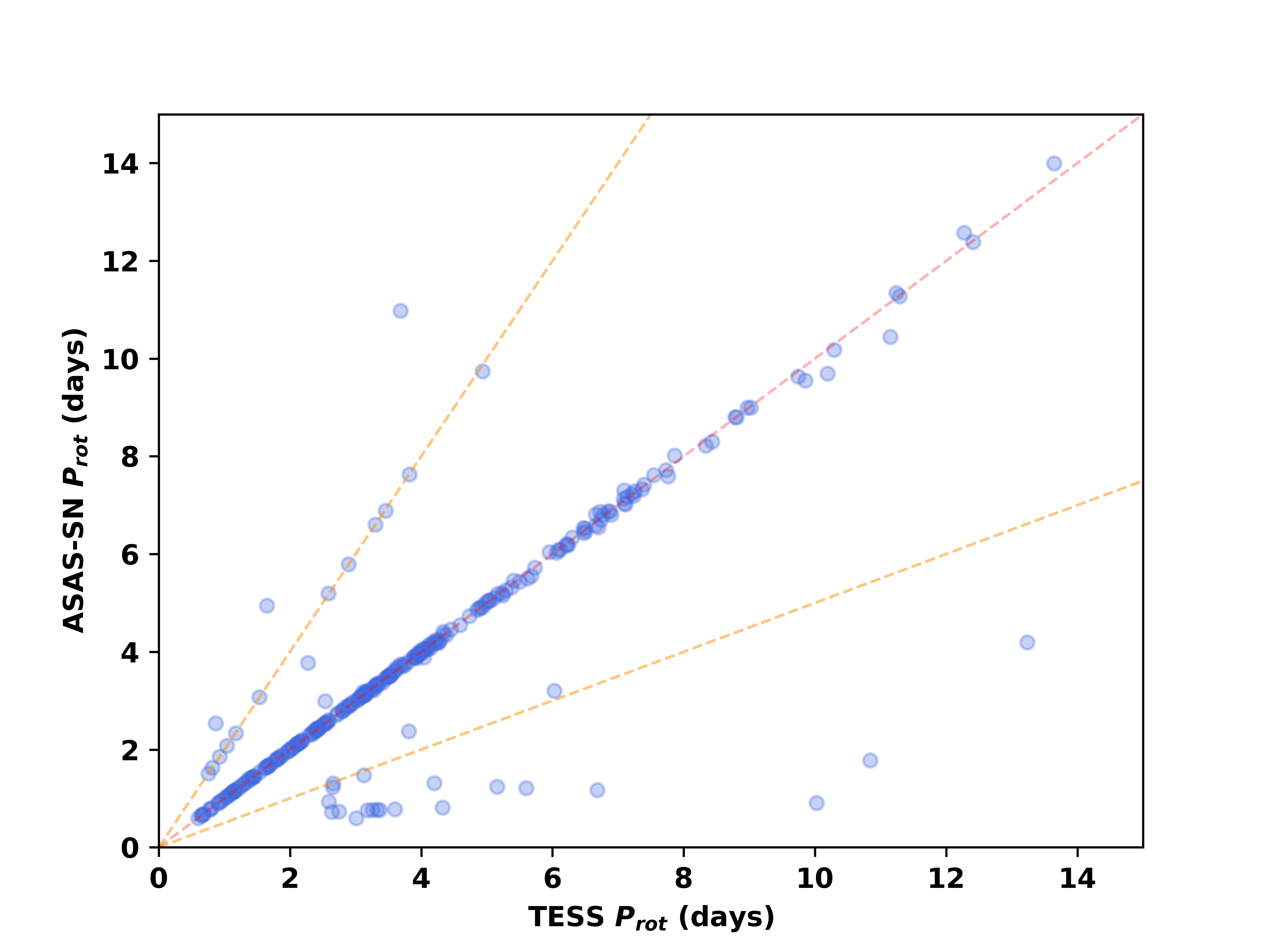}
    \caption{Left panel: TESS periods versus periods from the literature. Right Panel: TESS Periods versus ASAS-SN periods \citep{Jayasinghe_2018,Jayasinghe_2019}. Error bars for TESS periods are smaller than the marker size. Dotted lines represent 1:2, 1:1, and 1:1/2 relationships, respectively.}
    \label{fig:protcomparison}
\end{figure}

\noindent We compile stellar rotation periods from some multiepoch and multiband studies performed in selected regions of the OSFC. Table \ref{tab:compiledperiods} lists the studied regions with the respective nominal stellar ages, the analysis method used to obtain the rotational period, and the photometric band used for each selected reference. We have included rotational periods obtained from the All-Sky Automated Survey for Supernovae (ASAS-SN) database that provides $\sim$ 4 years baseline LCs for sources brighter than V $\lesssim$ 17 \citep{Jayasinghe_2018, Jayasinghe_2019}. Figure \ref{fig:protcomparison} shows comparisons between the compiled periods and the TESS periods derived in this work regardless of different databases for the LCs (e.g., cadence, photometric bands, observational windows, range of brightness, sensitivity, and technique). 
We found that 80\% of the sources have TESS periods that differ by less than 5\% from the literature values.

\begin{table}[H]
\caption{Compilation of rotational periods in the OSFC.}
\label{tab:compiledperiods}
	\begin{center}
	\begin{tabular}{lllll}
		\hline
		\hline
		Region & Age & Filter Band & $\#$ stars & Reference \\
		 & (Myr) &  &  \\
		\hline
		ONC & 1-2 & $I_{c}$ & 254 & \citet{Stassun1999} \\
		ONC & 1-2 & \textit{J}, \textit{H}, \textit{K} & 233 & \citet{Carpenter2001} \\
		ONC & 1-2 & $I_{c}$ & 138 \tablenotemark{a} & \citet{Rebull2001,Rebull2006a} \\
		ONC & 1-2 & [3.6], [4.5], $I_{c}$, \textit{J}, \textit{K} & 150 & \citet{Morales_Calderon_2011} \\
		ONC & 1-2 & ESO filter 851$\sim I_{c}$ & 369 & \citet{Herbst_2002} \\
		ONC & 1-2 & \textit{V, R, I, J, H, K} & 29 & \citet{Frasca_2009} \\
		ONC & 1-2 & \textit{V, R, I} & 148 & \citet{Parihar_2009} \\
		$\sigma$ Ori & 3 & $I_{c}$ & 84 & \citet{Cody2010} \\
		Orion OB1 Association & 4-10 & \textit{V}, $R_{c}$, $I_{c}$ & 564 & \citet{Karim2016} \\
		OSFC & ... & \textit{B, V} & 40 & \citet{Marilli_2007} \\
		OSFC & ... & \textit{V} & $\sim$1000 & \citet{Jayasinghe_2018, Jayasinghe_2019}\\
		\hline
	\end{tabular}
	\end{center}
	\tablenotetext{a}{Sources of \citet{Rebull2001} with periods reported in \citet{Rebull2006a}}
\end{table}

\subsection{Reddening, masses, and ages}
\label{sec:massage}

\noindent We use the \textit{MassAge} code (Hernandez et al. in preparation) to obtain reddening, luminosities, stellar ages, and masses for each kinematic member (with T$_{eff}<$6500 K) of the OSFC. The \textit{MassAge} code uses the effective temperature from the APOGEE-net pipeline (\S \ref{sec:vsini}), photometry (Gp, Rp, and Bp), and parallax from Gaia-EDR3 \citep{GAIA_EDR3}, and the J and H magnitudes from 2MASS \citep{2MASS2003}\footnote{Since K-band photometry could be contaminated by emission from disks on the OSFC sources, it was not included in this procedure}. The uncertainties in the estimated values are obtained using the Monte Carlo method of error propagation \citep{Anderson1976}, assuming Gaussian distribution for the uncertainties in the input parameters to generate 500 artificial points per each source. The adopted final results and uncertainties of the estimated values are the median and the median absolute deviation (MAD; 1$\sigma$=1.4826$\times$MAD), respectively.\\

\noindent To estimate the reddening to each star of our sample, the \textit{MassAge} code tests several values of A$_V$ until we obtain the best match between the observed colors ( [M$_{\lambda i}$ - M$_0$]$_{obs}$ ) and the standard colors affected by reddening ( [M$_{\lambda i}$ - M$_0$]$_{std}$ + [A$_{\lambda i}$/A$_V$-A$_0$/A$_V$]$\times$A$_V$ ), where M$_{\lambda i}$ are the Gaia/2MASS magnitudes (Bp, Rp, G, J, H) and A$_{\lambda i}$ is the extinction in magnitudes in those photometric bands. The standard colors were taken from \citet{Luhman2020} and the adopted reddening law A$_{\lambda i}$/A$_V$ values from \citet{Fitzpatrick2019}, assuming a canonical interstellar reddening law (R$_V$=3.1). 
The effective wavelength for each filter ($\lambda i$) was calculated using the filter transmission ($T_\lambda$) and the stellar flux ($S_\lambda$) in the respective bandpass as in equation \ref{eq:lfilter} \citep[e.g.,][]{Brown2016}.
The stellar flux was approximated using a PHOENIX synthetic reddened spectrum \citep{Husser2013} with a $T_{eff}$ similar to the target star. The reddening effect on the synthetic spectra correspond to the values of A$_V$ that we tested and the extinction law comes from \citet{Fitzpatrick2019}.\\

\begin{equation}
\label{eq:lfilter}
\lambda i=\frac{\int{\lambda T_\lambda S_\lambda d\lambda}}{\int{T_\lambda S_\lambda d\lambda}}
\end{equation}

\begin{figure}[htp]
    \centering
    \includegraphics[width=0.42\textwidth]{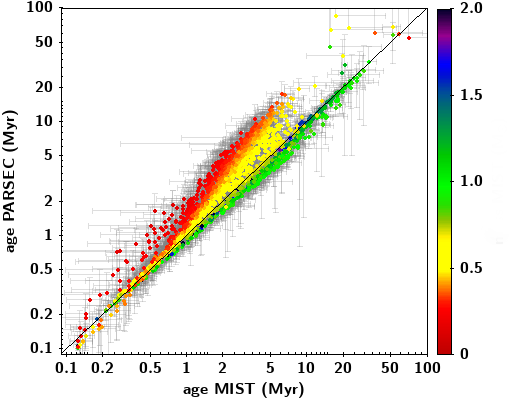}
    \includegraphics[width=0.42\textwidth]{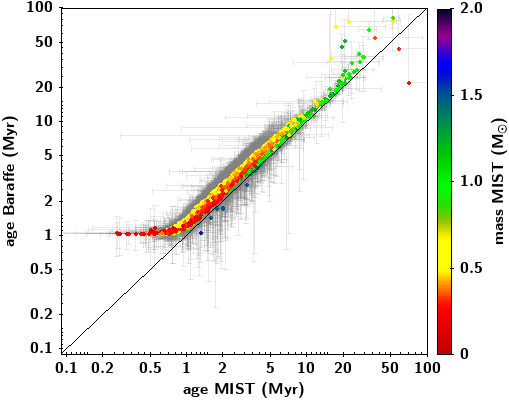}
    \caption{Comparison between the stellar ages derived by the MassAge code using different evolutionary models. Left panel:  ages derived using PARSEC-COLIBRI \citep{Marigo2017} versus MIST \citep{Dotter2016}. Right Panel: ages derived using \citet{Baraffe_2015} and MIST \citep{Dotter2016}. The color bar represent masses derived using MIST models.}
    \label{fig:ages}
\end{figure}

\noindent Once each star of our sample is located in the HR diagram, we estimated its stellar age and mass using three different evolutionary models: \citet{Baraffe_2015}, \citet[MIST;][]{Dotter2016}, and \citet[PARSEC-COLIBRI;][]{Marigo2017}. Note that the adopted luminosities were derived from the reddening-corrected \textit{J} band photometry, with the \textit{J} band bolometric correction for PMS stars from \citet{Pecaut2013}, and the corrected Gaia-EDR3 parallaxes. For each of the 500 artificial points, we selected the stellar mass and the stellar age corresponding to the closest theoretical point in the evolutionary model grid. Table \ref{tab:massage} includes the stellar parameters derived for the kinematic sample. It is important to mention that masses and ages derived using this method can be affected by unresolved binaries or stellar variability.  Figure \ref{fig:ages} shows comparisons between the stellar ages derived by the \textit{MassAge} code using the different evolutionary models. Stellar ages from MIST are in good agreement with those derived from \citet{Baraffe_2015}. In contrast, stars with stellar masses $\sim$0.5 $M_{\sun}$ or smaller have stellar ages from PARSEC-COLIBRI models systematically larger than those derived from MIST.

\begin{longrotatetable}
\begin{deluxetable*}{cccccccccc}
\tablecaption{Masses and ages for kinematic members with T$_{eff}<6500K$ \label{tab:massage}}
\tablehead{
\colhead{2massID}   &  \colhead{$T_{eff}$ (K)} & \colhead{L$_*$}      &\colhead{Av} & \colhead{Mass\tablenotemark{a}} &\colhead{Age\tablenotemark{a}} & \colhead{Mass\tablenotemark{b}} &\colhead{Age\tablenotemark{b}} & \colhead{Mass\tablenotemark{c}} &\colhead{Age\tablenotemark{c}}\\
\colhead{       }   &  \colhead{[K]}   & \colhead{[L$_{\odot}$]} &\colhead{[mag]} & \colhead{[M$_{\odot}$]} &\colhead{[Myr]} & \colhead{[M$_{\odot}$]} &\colhead{[Myr]} & \colhead{[M$_{\odot}$]} &\colhead{[Myr]}
}
\startdata
  05365409-0253155  &    4335.9 $\pm$    51.3  &      1.08 $\pm$    1.04  &	   0.36 $\pm$   0.07	&    0.94 $\pm$  0.06  &    1.98 $\pm$  0.32  &    0.81 $\pm$  0.05  &    1.55 $\pm$  0.24  &    1.04 $\pm$  0.06  &    2.27 $\pm$  0.40 \\ 
  05391151-0231065  &    3755.3 $\pm$    48.6  &      0.52 $\pm$    1.05  &	   0.80 $\pm$   0.11	&    0.50 $\pm$  0.04  &    1.30 $\pm$  0.21  &    0.56 $\pm$  0.04  &    1.53 $\pm$  0.28  &    0.62 $\pm$  0.05  &    1.75 $\pm$  0.41 \\ 
  05355077-0516291  &    4135.2 $\pm$    49.8  &      0.97 $\pm$    1.05  &	   0.85 $\pm$   0.11	&    0.75 $\pm$  0.05  &    1.37 $\pm$  0.23  &    0.71 $\pm$  0.02  &    1.26 $\pm$  0.15  &    0.84 $\pm$  0.04  &    1.57 $\pm$  0.26 \\ 
  05415845-0115486  &    3945.3 $\pm$    50.3  &      0.86 $\pm$    1.06  &	   3.43 $\pm$   0.20	&    0.59 $\pm$  0.04  &    0.98 $\pm$  0.17  &    0.60 $\pm$  0.04  &    1.01 $\pm$  0.17  &    0.74 $\pm$  0.03  &    1.24 $\pm$  0.20 \\ 
  05392147-0723300  &    3509.5 $\pm$    51.4  &      0.78 $\pm$    1.09  &	   4.84 $\pm$   0.24	&    0.33 $\pm$  0.03  &    0.38 $\pm$  0.14  &    0.30 $\pm$  0.03  &    0.36 $\pm$  0.16  &  \nodata $\pm$\nodata  &  \nodata $\pm$\nodata \\ 
  05365014-0641292  &    4039.1 $\pm$    47.3  &      0.95 $\pm$    1.07  &	   3.52 $\pm$   0.17	&    0.66 $\pm$  0.04  &    1.08 $\pm$  0.19  &    0.66 $\pm$  0.04  &    1.09 $\pm$  0.18  &    0.78 $\pm$  0.03  &    1.25 $\pm$  0.21 \\ 
  05320778-0655368  &    3540.2 $\pm$    51.7  &      0.21 $\pm$    1.08  &	   0.00 $\pm$   0.16	&    0.39 $\pm$  0.04  &    2.81 $\pm$  0.63  &    0.58 $\pm$  0.04  &    5.03 $\pm$  1.12  &    0.44 $\pm$  0.05  &    3.55 $\pm$  0.91 \\ 
  05392519-0238220  &    3917.5 $\pm$    50.0  &      1.08 $\pm$    1.06  &	   1.39 $\pm$   0.13	&    0.55 $\pm$  0.04  &    0.64 $\pm$  0.09  &    0.54 $\pm$  0.04  &    0.67 $\pm$  0.09  &  \nodata $\pm$\nodata  &  \nodata $\pm$\nodata \\ 
  05353907-0508564  &    4252.3 $\pm$    92.9  &      1.57 $\pm$    1.05  &	   0.96 $\pm$   0.10	&    0.80 $\pm$  0.09  &    0.88 $\pm$  0.23  &    0.73 $\pm$  0.05  &    0.78 $\pm$  0.14  &    0.97 $\pm$  0.06  &    1.09 $\pm$  0.16 \\ 
  05314751+0217129  &    5954.1 $\pm$    52.1  &      3.83 $\pm$    1.04  &	   0.00 $\pm$   0.15	&    1.43 $\pm$  0.04  &   10.53 $\pm$  0.97  &    1.47 $\pm$  0.04  &    9.85 $\pm$  0.87  &  \nodata $\pm$\nodata  &  \nodata $\pm$\nodata \\ 
\enddata 
\tablenotetext{a}{MIST}
\tablenotetext{b}{PARSEC}
\tablenotetext{c}{Baraffe+2015}
\tablecomments{Only a portion of the table is shown here. The full version is available in electronic form.}
\end{deluxetable*}
\end{longrotatetable}

\section{Results and Discussion}
\label{sec:results}

\subsection{$v\sin(i)$ vs $P_{\rm rot}$}
\label{sec:vsinper}

\begin{figure}[htp]
    \centering
    \includegraphics[width=0.5\textwidth]{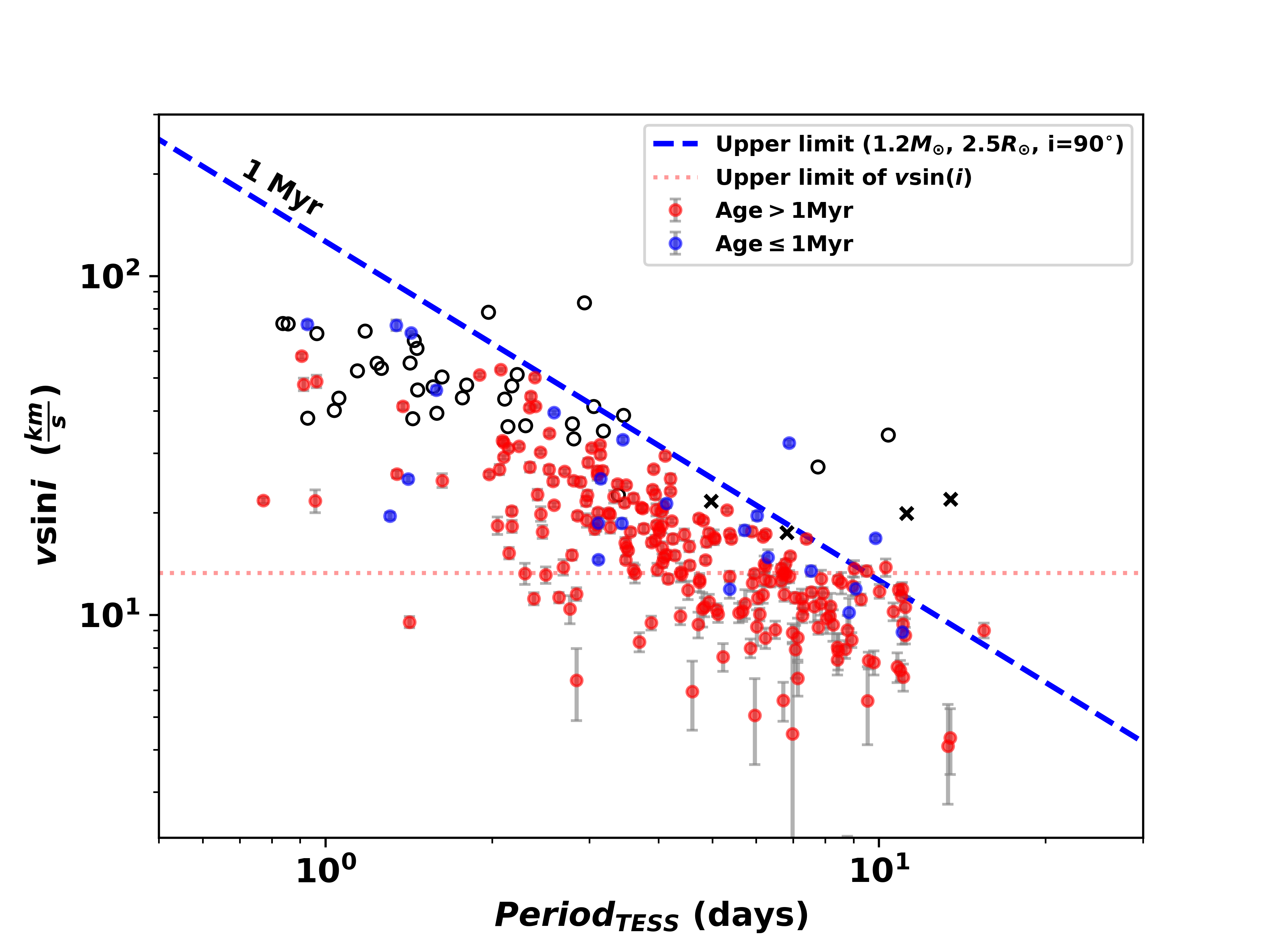}
    \caption{$v\sin(i)$ versus $P_{\rm rot}$ derived in this work. The blue dashed line shows a solid body model of $1.2M_{\odot}$, $2.5R_{\odot}$ at 1 Myr, $\sin i=1$. Black crosses are double-line spectroscopic binaries (SB2) and black open circles are uncertain SB2 \citep{Kounkel_2019}. Red dotted line show the upper limit to $v\sin(i)$=13.3 km~s$^{-1}$. The period error bars are smaller than the marker size.}
    \label{fig:velpertess}
\end{figure}

\noindent In Figure \ref{fig:velpertess} we plot the rotational period ($P_{\rm rot}$) obtained from our TESS LC analysis versus the $v\sin(i)$ parameter measured for 277 stars that are OSFC kinematic members. Assuming a solid body rotation, these parameters are related to the stellar radius ($R$) as follows:

\begin{equation}
\label{eq:v_Prot}
v\sin(i)=\frac{2\pi R\sin(i)}{P_{\rm rot}}
\end{equation}

\noindent As reference, we also plot the theoretical upper limit of $v\sin (i)$ for each $P_{\rm rot}$ assuming a maximum stellar radius of 2.5~$R_{\odot}$ and a rotation axis inclination of i=90$\degr$. Using solar metallicity MIST models \citep{Dotter2016}, this corresponds to a maximum stellar mass of 1.2 $M_{\odot}$ with an age of 1 Myr, that of the youngest sub-associations in the OSFC. 

\noindent We have considered $v\sin(i)$ measurements for double-line spectroscopic binaries (SB2) in Figure \ref{fig:velpertess} reported by \citet{Kounkel_2019}. As expected, $v\sin(i)$ estimations could be strongly affected by stellar companions. In this sense, Figure \ref{fig:velpertess} can be used as a diagnostic plot to detect potential candidates for spectroscopic binaries. To inspect possible binaries in our sample, we have added stars that did not pass the binarity criteria of the \citet{Kounkel_2019} analysis, reported as unresolved SB2 or star dominated by starspots (SB2/sp). In the latter case, one does not expect a $v\sin(i)$ larger than the one derived from the rotation period and the typical radius at 1 Myr (blue dashed line in Fig. 6). Generally, we expect binary stars or stars younger than 1 Myr to be located above the blue dashed line. 
Two of the stars above the blue dashed line have ages younger than 1 Myr (\S \ref{sec:massage}), and eight stars results possible SB2, an spectroscopic follow up is required to confirm the nature of these targets.\\

\noindent An anti-correlation between $v\sin(i)$ and $P_{\rm rot}$ is expected according to equation \ref{eq:v_Prot}. The scatter in this trend is mainly produced by different spin axis inclinations (i) and different stellar radii in the sample. Undetected stellar companions could also affect the anti-correlation of Figure \ref{fig:velpertess}.  There is a possibility that broader spectral features are produced by unresolved combined features from both components in binary stars instead of those produced by fast rotation \citep[e.g., ][]{Maxted2008,Hernandez_2014}. 
Additional scatter can be produced by latitudinal differential rotation, which increases toward smaller inclination angles \citep{Reiners2002}.
However, \citet{Barnes2005} show that differential rotation decreases with temperature. In particular, differential rotation is low for stars with $T_{eff}<$5000K and disappears at 3500K. Since most stars in our sample have $T_{eff}<$5000K, we expect this effect to be minimal in our sample.
In addition, the inhibition of convective transport of energy by stellar magnetic fields and/or the presence of cool magnetic starspots can produce the so-called radius inflation effect \citep{Jackson2014,Lanzafame2017,Somers2020}. This effect is expected to be below 15\% \citep{Kesseli2018}. Regardless of these effects, plots like Figure  \ref{fig:velpertess} can be used as a tool for searching binary candidates, as these would fall above the reference line.\\

\noindent Combining $v\sin(i)$ measurements (\S \ref{sec:vsini}), the TESS rotational periods (\S \ref{sec:tessperiods}), and the $L_{\ast}$ and $T_{eff}$ estimations (\S \ref{sec:massage}), we derived the ratio of projected radius to the model radius ($\frac{R\sin(i)}{R^{\prime}_{*}}$) as follows: 

\begin{equation}
\label{eq:sini}
 \frac{R\sin(i)}{R^{\prime}_{*}}=\frac{v\sin(i) P_{\rm rot}}{2\pi} \times \sqrt{\frac{4\pi\sigma}{L_\ast}}T_{eff}^2
\end{equation}

\noindent where the second term is the inverse of the model radius ($R^{\prime}_{\ast}$) derived from the luminosity and $T_{eff}$ according to the Stefan-Boltzmann law. In the subsequent analyses, the sample is restricted to those objects with $v\sin(i)>13$ km~s$^{-1}$ (the instrumental resolution limit).\\

\begin{figure}[htp]
    \centering
    \includegraphics[width=0.5\textwidth]{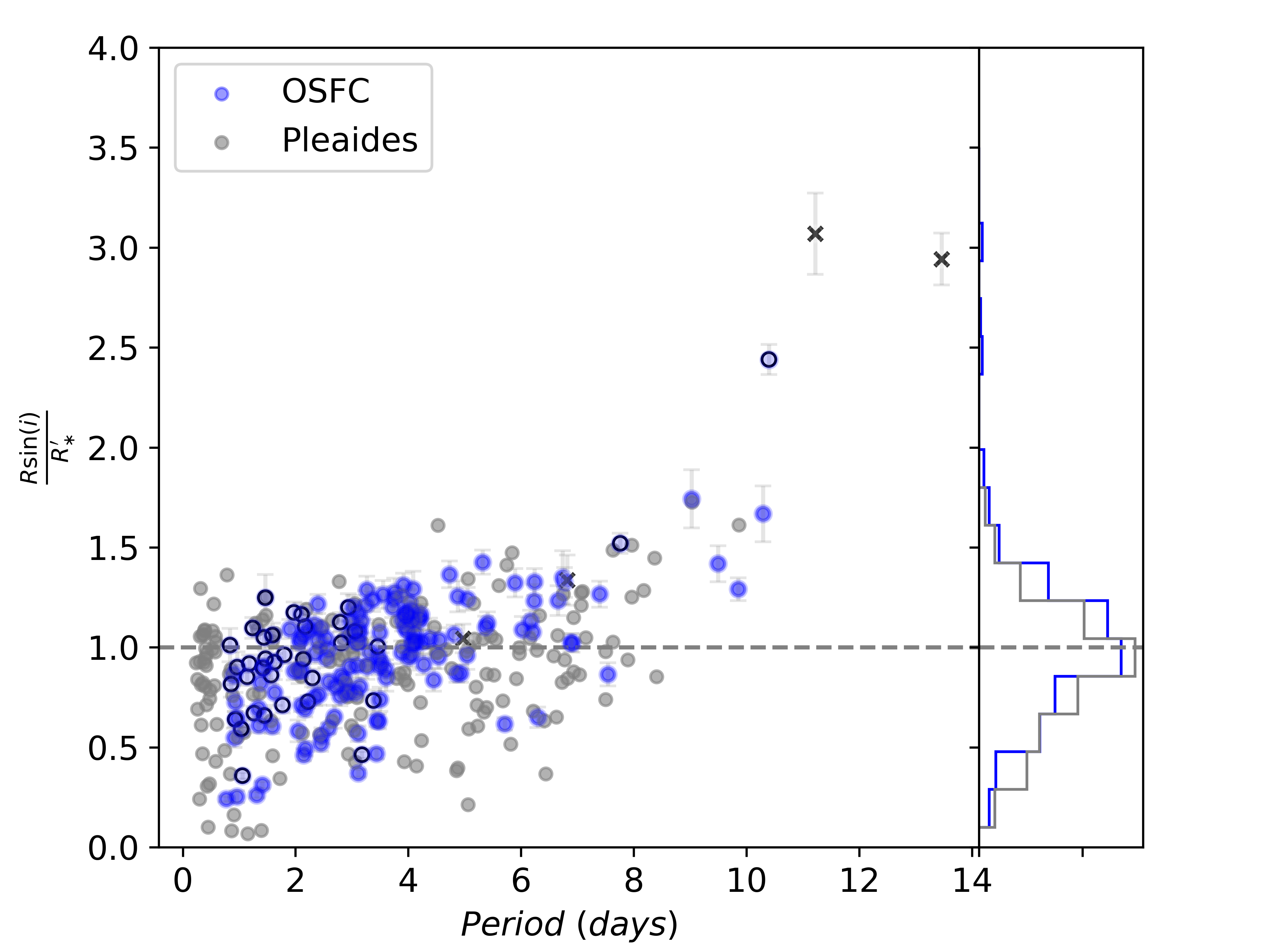}
    \caption{$\frac{R\sin(i)}{R^{\prime}_{\ast}}$ of the OSFC compared with the Pleiades
    both as a function of $P_{\rm rot}$. The left panel show the OSFC stars with $\frac{R\sin(i)}{R^{\prime}_{\ast}}$ and periods estimated in this work (blue dots). The grey dots represent the Pleiades stars with $\frac{R\sin(i)}{R^{\prime}_{\ast}}$, and periods compiled from \citep{Hartman_2010}. Black crosses are double-line spectroscopic binaries (SB2) and black open circles are uncertain SB2 or spotted star \citep{Kounkel_2019}. The dashed line represent the mathematical limit for the inclination ($\sin(i)=1$). Right panel: Normalized histogram of the vertical axis.}
    \label{fig:sini}
\end{figure}

\noindent We found that stars trace a sequence on the $\frac{R\sin(i)}{R^{\prime}_{*}}$ versus  $P_{\rm rot}$ diagram, with $\frac{R\sin(i)}{R^{\prime}_{*}}$ tending to increase with $P_{\rm rot}$. A similar trend was reported before for CTTS and WTTS \citep[e.g.,][]{ Artemenko2012,Venuti2017}. We include in Figure \ref{fig:sini} stars of the Pleiades cluster \citep[age=125 Myr; e.g.,][]{Somers2017}, using the measurements of $v\sin(i)$, $P_{\rm rot}$, and stellar radii reported by \citet{Hartman_2010}. It is apparent that stars in Pleiades and stars in the OSFC share similar behavior in Figure \ref{fig:sini}.  There are a significant number of stars showing a larger projected rotational radius than the expected theoretical radius. This radius inflation effect has been observed previously and has been attributed to enhanced magnetic activity and the presence of stellar spots \citep[e.g.][]{Somers2015,Somers2020, Bouvier2016,Jackson_2016,Lanzafame2017,Jaehnig2019}. Other effects can increase the $\frac{R\sin(i)}{R^{\prime}_{*}}$ parameter. e.g., systematic offsets of $T_{eff}$ and $L_\ast$  due to starspots, plages, or contamination by binaries \citep{Somers2017}. Also, systematic errors in the estimation of $v\sin(i)$ can be introduced by unresolved binaries \citep{Kesseli2018} or by using a not suitable limb darkening law in pre-main-sequence stars \citep{Rhode2001}. 
Finally, It is challenging to detect photometric rotational modulation for stars with a low inclination, especially for those with longer periods. Low values of $\frac{R\sin(i)}{R^{\prime}_{*}}$ are frequent at low $P_{\rm rot}$ and diminish with the increase of $P_{\rm rot}$.
Exploring the origin of stars with $\frac{R\sin(i)}{R^{\prime}_{*}}>1$ is beyond the scope of this paper. A detailed analysis of the radius inflation effect and the derived inclination pole  distributions will be addressed in future works.

\subsection{Rotational Evolution}
\label{sec:rotevol}

\noindent We use $v\sin(i)$ and individual stellar ages of the kinematic members in the OSFC sample (\S \ref{sec:members2}) to examine evolutionary trends during the first million years of their life. We consider $v\sin(i)$ measurements that have fractional errors smaller than 10$\%$. Additionally, stars identified as multiple stellar systems \citep{Kounkel_2019} were excluded from our sample. We prefer the use of $v\sin(i)$ instead of rotational periods because it is difficult to obtain rotational periods from TESS LCs in clustered young stellar regions (e.g., $\sigma$ Ori, ONC). Moreover, the complex brightness variability in stars with accreting disks prevents obtaining estimations of rotational periods in most CTTS. \\

\begin{figure}[htp]
    \centering
    \includegraphics[width=0.45\textwidth]{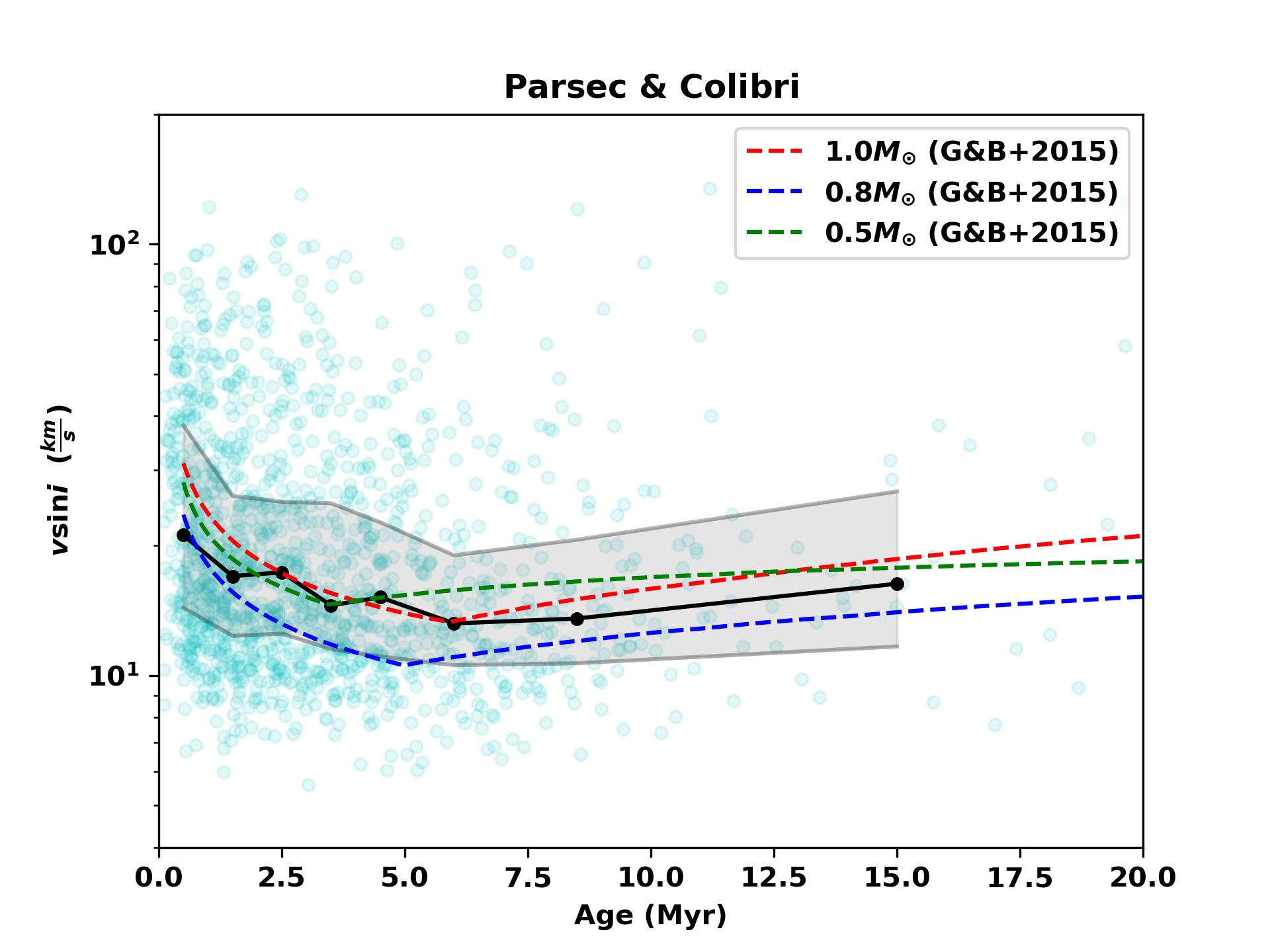}
    \includegraphics[width=0.45\textwidth]{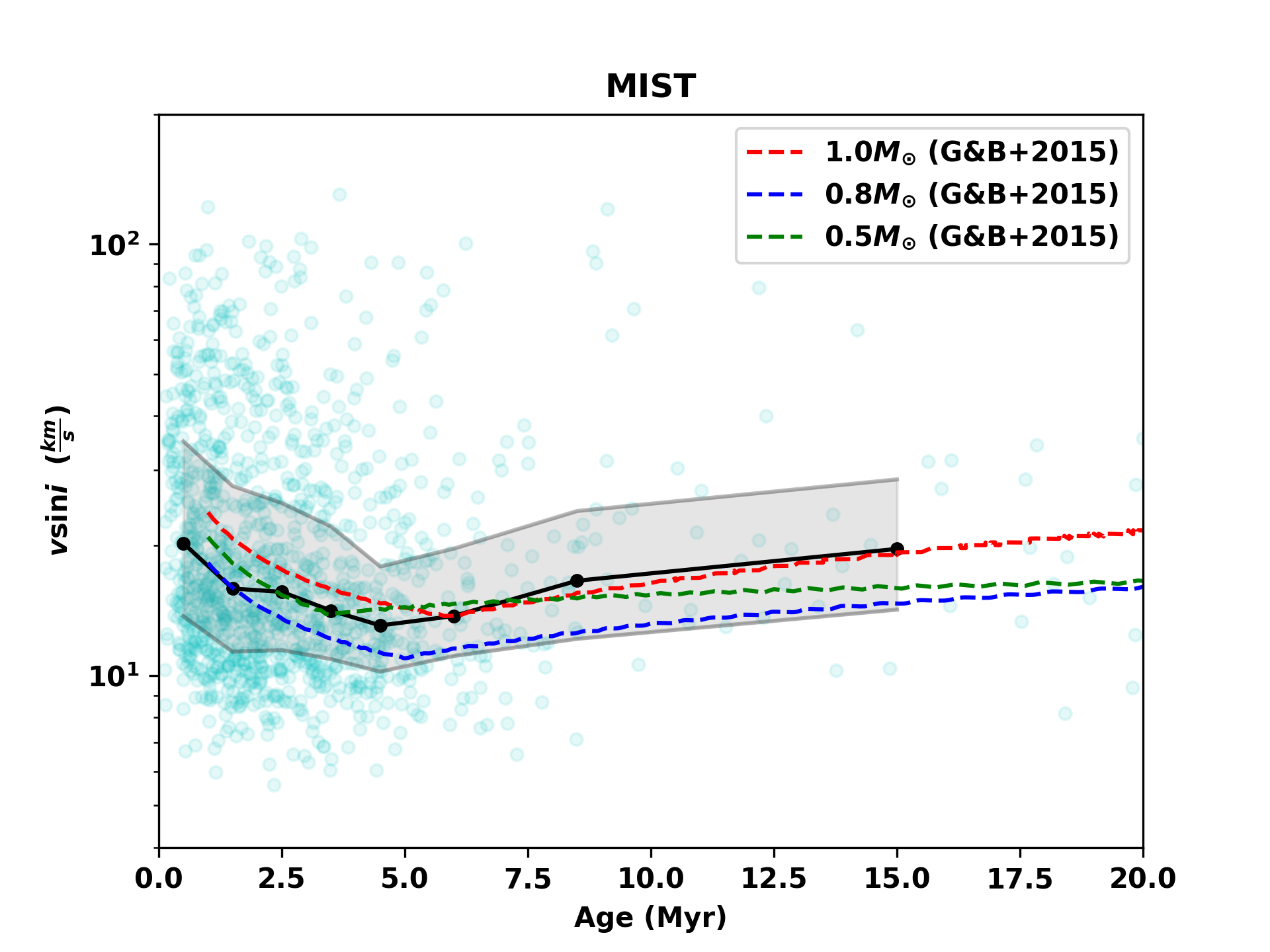}
    \includegraphics[width=0.45\textwidth]{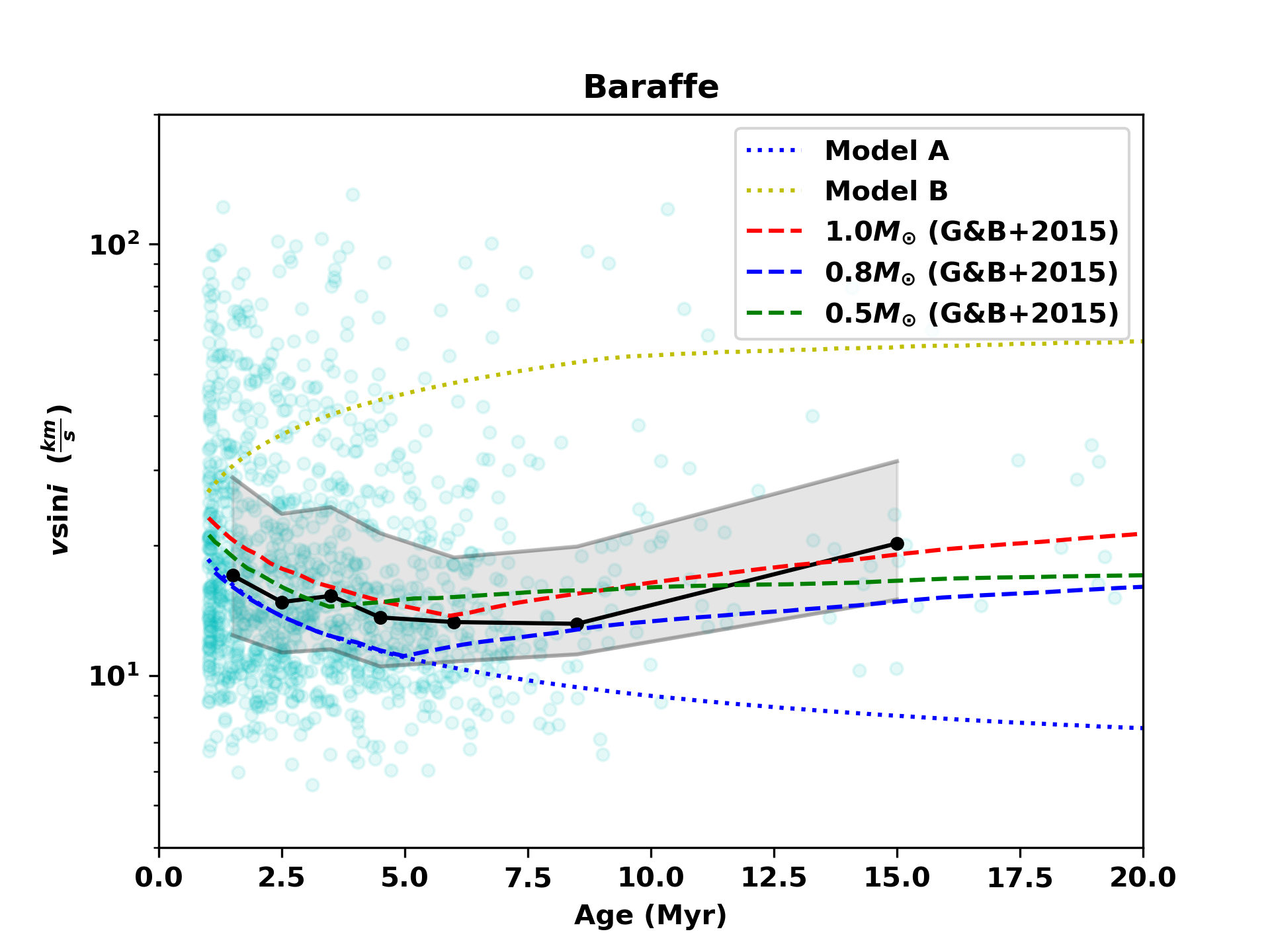}
    \caption{Evolution of $v\sin(i)$ versus age derived using the evolutionary models by PARSEC-COLIBRI \citep{Marigo2017}, MIST \citep{Dotter2016}, and Baraffe \citep{Baraffe_2015} for the sample of Orion's kinematic members (aqua green dots). The median of all points is represented by the black points joined by solid lines, while the gray zones define the interquartile range. Model A (disk-locking model up to 20 Myr) is represented by the blue dotted line. Model B (diskless star model) is shown as the yellow dotted line. Results from \citet{Gallet_2015} models are transformed to rotational velocities, assuming solid body rotation, $\sin i=1$, and radius evolution from the respective evolutionary models. These are represented with dashed lines. Model A coincides with 0.8 $M_{\odot}$ \citep{Gallet_2015} model during the first 5 Myr. Model A and B are shown only in the third plot for illustrative purposes.}
    \label{fig:trendrotation}
\end{figure}

\noindent Figure \ref{fig:trendrotation} shows the evolution of $v\sin(i)$, considering ages derived with three different evolutionary models \citep{Baraffe_2015,Dotter2016,Marigo2017} as described in \S \ref{sec:massage}.
We find a large scatter of $v\sin(i)$ at all ages. This is caused by several factors \citep[e.g.][]{Stassun1999}, among them, variations on inclination angles ($\sin(i)$), on stellar masses, uncertainties in the determination of ages, and differing initial conditions of angular momentum evolution (e.g. the distribution of initial rotational periods).\\
To investigate the general evolution of stellar rotation for the kinematic members, we plot the median $v\sin(i)$ values (solid line) and the interquartile range (IQR) estimated for different age bins; The IQR corresponds to 50$\%$ central confidence interval (gray zones).
We select the age bins to contain approximately the same number of stars per bin (0$\rightarrow$1, 1$\rightarrow$2, 2$\rightarrow$3, 3$\rightarrow$4, 4$\rightarrow$5, 5$\rightarrow$7, 7$\rightarrow$10, y 10$\rightarrow$20 Myr).\\

\noindent \citet{Gallet_2015} present parametric spin-evolution models ranging from 1 Myr to 10 Gyr for slow, medium, and fast rotators for three selected stellar masses (0.5, 0.8, and 1.0 $M_{\odot}$). These models include star-disk interaction, a magnetized stellar wind, core-envelope decoupling, and structural stellar evolution (i.e., mass, radius, moment of inertia) of the inner radiative core, and the outer convective envelope provided by \citet{Baraffe1998}. During the accreting disk lifetime, the angular velocity for the stellar surface is constant as a consequence of the balance between the angular momentum added to the star from disk accretion and that extracted from it through the magnetic star-disk interaction \citep{Matt2012}. After the end of the accretion phase, the angular velocity evolution for PMS low-mass stars is regulated mainly by the structural stellar evolution and magnetized stellar winds, once the star arrives to the main sequence \citep{Kawaler1988}. By transforming angular velocities ($\omega$) into rotational velocities ($V_{rot} = R_{\ast}\omega$), we plot in Figure \ref{fig:trendrotation} the prediction of the models for medium rotators \citep{Gallet_2015}. Each model is associated to a stellar mass of 0.5, 0.8, and 1.0 $M_{\odot}$, that in turn is related to a disk lifetime, 3, 5, and 6 Myr, respectively. Regardless of the stellar evolution models used for estimating stellar ages, the general rotational behavior of the kinematic members follows the general trend predicted by these models reasonably well.\\

\noindent For reference, we over-plot in the upper-left panel of Figure \ref{fig:trendrotation}, a couple of "toy models" showing extreme scenarios for angular momentum evolution (dotted lines). Model "A" describes a star with an accreting disk. The disk-locking effect keeps a constant angular velocity up to 20 Myr. We assume a solid body with a constant period of 6 days, similar to that adopted by \citet{Gallet_2015} in their 0.8 $M_{\odot}$ model. The evolution of $v\sin(i)$ is proportional to the stellar radius ($v_{rot}=\frac{2\pi R_{\ast}(t)}{P_{\rm rot}}$), which is reduced by 50\% between 1-10 Myr \citep{Baraffe_2015}. In contrast, the model "B" describes a diskless star. The $v\sin(i)$ evolution is derived only from the stellar radius evolution and the AM conservation. As a starting point, we use the AM ($J_0$) derived from the empirical model of \citet{Kawaler_1987} and referred in \citet[Equation (1)]{Landin_2016}:

\[J_{0}=I\omega=1.566\times 10^{50}\bigg(\frac{M_{\ast}}{M_{\odot}}\bigg)^{0.985}\,\,[cgs].\]

\noindent We consider a solid body rotator with a moment of inertia $I=k^{2}M_{\ast}R^{2}_{\ast}$; The gyration radius ($k$) and the stellar radius ($R_{\ast}$) are obtained from the stellar evolutionary models of \citet{Baraffe_2015}.
In this case, the $v\sin(i)$ evolution is inversely proportional to the stellar radius ($v_{rot}=\frac{J_{0}}{k^{2}(t)M_{\ast}R_{\ast}(t)}$).\\

\noindent In agreement with \citet{Gallet_2015}, the $v\sin(i)$ statistical trend observed in the kinematic members of the OSFC (see Figure \ref{fig:trendrotation}) suggests the existence of two phases in the early PMS evolution. The transition time between these phases is similar to the inflection point observed in the evolution of the disk frequency, where the fraction of low-mass accretors decreases about 90\% during the first 5-6 Myr \citep{Hernandez2007, Fedele2010, Briceno_2019, Manzo_2020}. During the first phase (age $\lesssim$ 5-6 Myr), $v\sin(i)$ decreases in spite of the rapid contraction. The torques originated from the magnetic coupling between the star and the disk remove efficiently the angular momentum from the system. Although details about these mechanisms are still uncertain due to the lack of knowledge of the interplay between accretion and ejection phenomena.  
In the second phase (age $\gtrsim$ 5-6 Myr), the disk effect on the angular momentum evolution has stopped, causing the stellar rotation to be regulated by the structural stellar evolution and stellar winds.\\

\noindent The key takeaways from the present subsection are the following:
\begin{itemize}
    \item $v\sin(i)$ decreases statistically for the first 5-6 Myr of the star's life, which suggests the existence of mechanisms that extracts angular momentum from the stars. Likely, disk-locking is the most important of these mechanisms.
    \item Beyond 5-6 Myr, the disk disappears, and $v\sin(i)$ tends to increase due to stellar contraction and angular momentum conservation.
 
\end{itemize}

\subsection{Rotation and accretion}
\label{sec:rothalpha}

\begin{figure}[htp]
    \centering
    \includegraphics[width=0.5\textwidth]{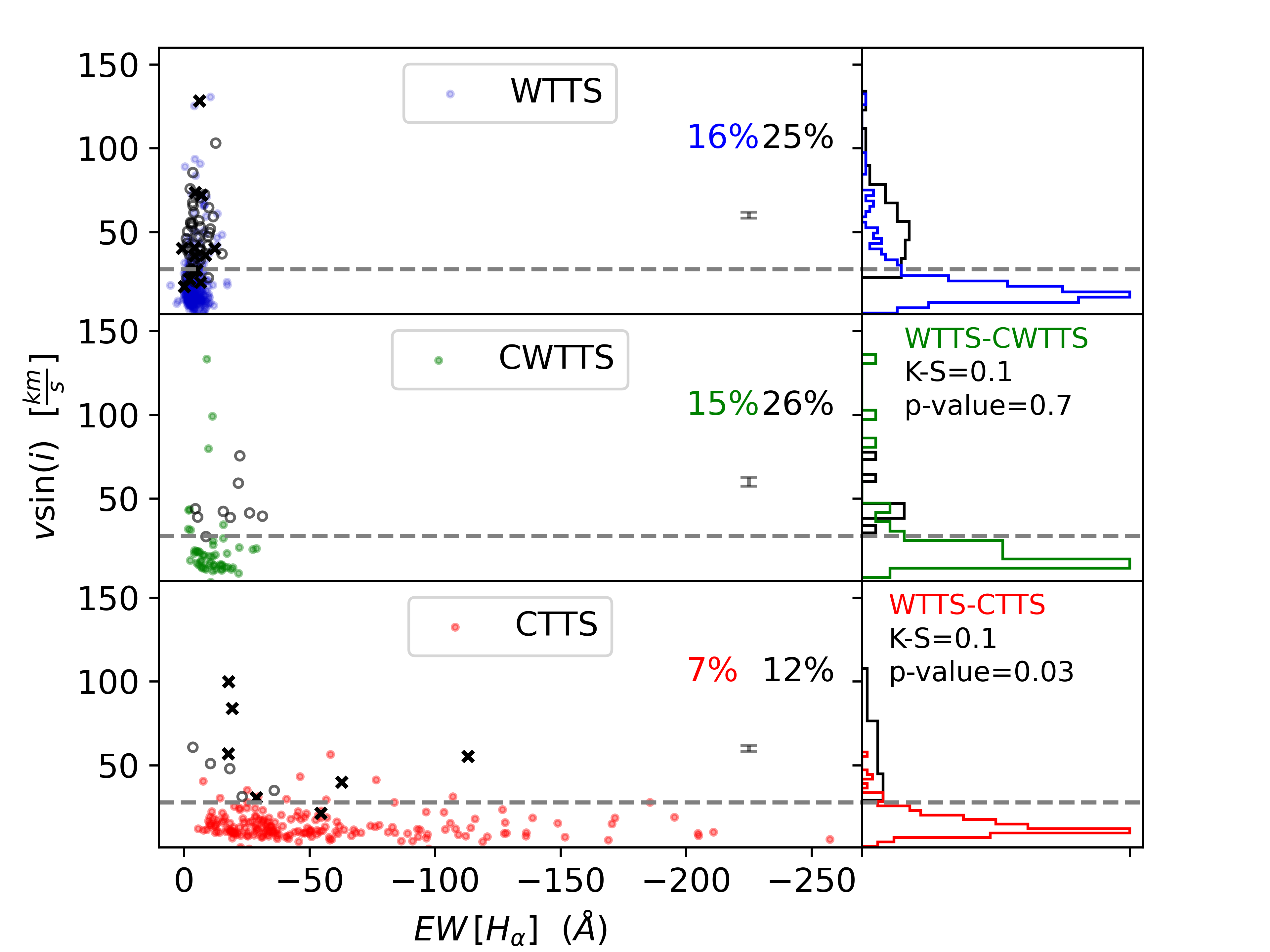}
    \caption{Left panels: $v\sin(i)$ versus $EW_{H_{\alpha}}$ for TTS. Black crosses are double-line spectroscopic binaries (SB2) and black open circles are uncertain SB2 or spotted star (SB2/sp) \citep{Kounkel_2019}. The dashed line split our slow rotators from the fast ones, noted as 28 km s$^{-1}$. The colored percentages show the fraction of single stars above the dashed line, and the black percentage shows the fraction for the whole sample (single, binaries and multiple systems) above dashed line. Right panels: histograms of $v\sin(i)$ for both the single stars and SB2 or SB2/Sp. The results of the K-S test between WTTS, CWTTS, and CTTS are displayed above histograms.}
    \label{fig:velhalpha}
\end{figure}

\noindent Using the classification of TTS derived from the equivalent width of $H\alpha$ (\S \ref{sec:members2}), we analyze the $v\sin(i)$ distributions of the CTTS, WTTS, and CWTTS samples. For all the samples of TTS (Figure \ref{fig:velhalpha}), we plot a statistical limit that separates the slow rotators from the fast rotators (gray dashed line), the limit is defined so that 25$\%$ of WTTS have $v\sin(i)$ greater than this limit (i.e, $v\sin(i)=28$ km~s$^{-1}$).

\noindent In general WTTS show a large range of stellar rotation velocities, such that 25\% (96/383) of the stars are considered fast rotators. In contrast, only 12\% (24/206) of the CTTS can be classified as fast rotators (lower panel of Figure \ref{fig:velhalpha}).
Excluding binary stars and binary candidates reported by \citet{Kounkel_2019}, the fractions of fast rotators are 16\% (52/330) and 7\% (13/194) for WTTS and CTTS samples, respectively. In general, single CTTS have $v\sin(i) \leq 50$ km~s$^{-1}$. \\

\noindent A visual inspection of the TTS distributions suggests that CTTS exhibit distinct rotation properties from WTTS.
A Kolmogorov-Smirnov (K-S) test applied to both populations, rejecting binary candidates, supports this idea of statistically distinct rotation properties for the two groups. Indeed, the test yields a probability of only 0.03 that the two distributions in $v\sin(i)$ corresponding to CTTS and WTTS are extracted from the same parent distribution (Figure \ref{fig:velhalpha}).
These results provide a clear indication of a statistical connection between disk and rotation properties across the OSFC.\\

\noindent \citet{Jayawardhana_2006} found similar results in the $\eta$ Cha and the TW Hydrae ($\sim6\text{-}8$ Myr) stellar groups. They report that all accretors are slow rotators ($v\sin(i)\leq$ 20 km~s$^{-1}$), while non-accretors cover a larger range in rotational velocities. These results suggest a clear signature of disk-braking on the CTTS sample.\\

\noindent In spite of small numbers of CWTTS stars in comparison to WTTS, our results do not show any statistical difference between CWTTS class and WTTS. From the p-value of 0.7 obtained from the K-S test, we are not allowed to discard the null hypothesis that the two populations are extracted from the same parent distribution.
The fast rotator fraction of CWTTS is 26\% (16/61) in comparison with the 25\% of the WTTS suggests that CWTTS are stars reaching the end of their accretion phase, and could be less affected by the disk-braking effect than accreting stars.\\

\begin{figure}[htp]
    \centering
    \includegraphics[width=0.5\textwidth]{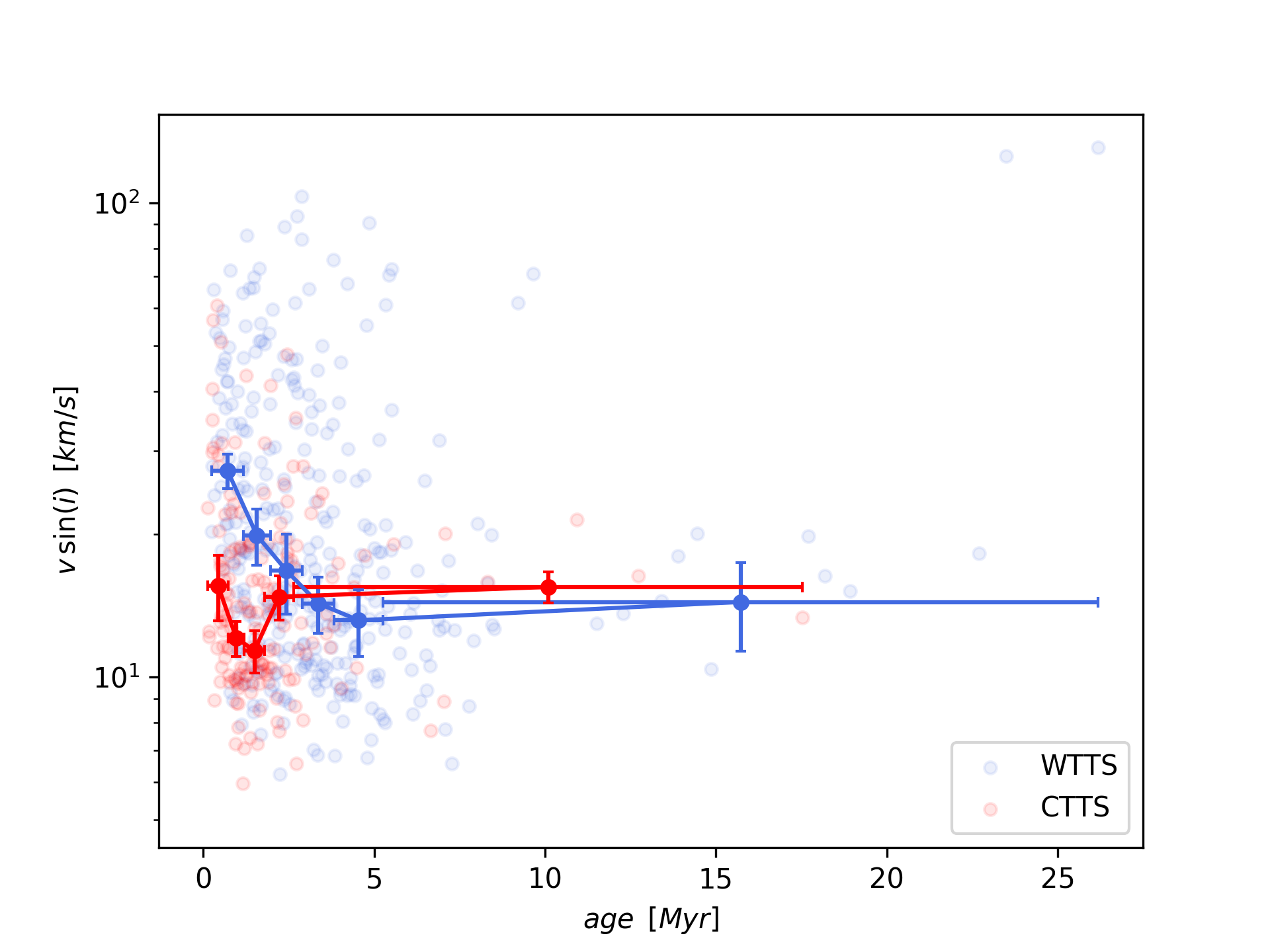}
    \caption{$v\sin(i)$ evolution for CTTS and WTTS in OSFC. Each solid dot represents the median per bin. The error bars are obtained considering Poissonian uncertainties. Differences between CTTS and WTTS are found to be significant only during the first $\sim$1.5 Myr in agreement to the disk-locking scenario.}
    \label{fig:veltts}
\end{figure}

\noindent Following the same methodology described in \S \ref{sec:rotevol}, we plot $v\sin(i)$ as a function of age for our sample of TTS (Figure \ref{fig:veltts}). 
The results suggest that, statistically, CTTS rotate slower than WTTS during the first 1.5 Myr, when the separations in $v\sin(i)$ between these two classes are larger than the Poisson uncertainties $\frac{\sigma_{v\sin(i)}}{\sqrt{N}}$. 
This result also suggests that the disk locking scenario applies for TTS in the OSFC. In addition, the decrease in the rotation rate of WTTS suggests that, although stars have recently lost their inner disk, they have not had enough time to increase their rotation, as suggested by other studies \citep{Jayawardhana_2006,Biazzo_2009,Nguyen_2009}. After 2 Myr, the rotational velocities of CTTS are comparable with those observed in WTTS.
Other causes of angular momentum loss in WTTS such as the stellar winds are expected to play important roles in the angular momentum evolution. Theoretical modelling suggest that winds may be present even at low stellar accretion rates, specially in stars that are magnetically active with strong non-axisymmetric fields that promove angular momentum losses via flaring events \citep{Nicholson_2018}. The absence of a disk creates favorable conditions for wind formation with higher mass-loss rates that may spin-down the star in time scales of the same order than the CTTS-WTTS transition time \citep{Sauty2011}. Additional studies of confirmed TTS older than 4 Myr are needed to obtain a statistically robust comparison between CTTS and WTTS at the end of the disk evolution phase.\\

\noindent The key takeaways from the present subsection are the following:
\begin{itemize}
    \item In general, CTTS's have a smaller rotational velocity than WTTS supporting the scenario in which the accreting disk regulates the angular momentum of disk bearing-stars.  
    \item The distribution of $v\sin(i)$ of CWTTS resembles that of the WTTS instead of CTTS. This suggests that the rotation rate of these stars, which are nearly at the end of their accretion phase, could be less affected by the disk-braking effect.
\end{itemize}

\section{Summary and Conclusions}
\label{sec:summary}
\noindent We have presented an evolutionary study of the stellar rotation in the Orion star forming complex for a very robust sample of kinematical members and spectroscopically confirmed TTS to date. We processed and analyzed 3501 LCs from TESS observations along the OSFC. Throughout a visual inspection to each LC, we have classified $1157$ stars as periodic variables. We have studied the measurements of $v\sin(i)$ for 1935 stars with stellar masses and ages estimations using different evolutionary models.
\\

\noindent The main results of this work are summarized as follows:\\
\begin{enumerate}
    \item We found an empirical trend of stellar rotation $v\sin(i)$ with age. It clearly shows a transitional phase on the stellar rotation at 5-6 Myr. This result is consistent with the disk lifetime used by \citet{Gallet_2015} in their AM evolutionary models for stars with masses between 0.8$M_\sun$ and 1$M_\sun$, and in agreement with 
    the timescale for disk dissipation obtained by studies of protoplanetary disks \citep{Hernandez2007,Carpenter_2006}.
    \item Analysis of the $v\sin(i)$ distributions shows that CTTS are slow rotators suggesting that disk-braking effect regulates the angular momentum evolution of accreting stars, maintaining the rotation rates substantially below the ($v\sin(i)<28$ km~s$^{-1}$). In addition our data suggest that CTTS with $v\sin(i)>28$ km~s$^{-1}$ are likely binary systems. 
    \item Stars reaching the end of their accretion phase (CWTTS) have similar statistical properties than WTTS, suggesting that this sample could be less affected by the disk-braking effect. 
    
\end{enumerate}

\section*{acknowledgments}

\noindent A special thank to Florian Gallet who kindly shared his models with us. JS acknowledges from the CONACYT by fellowship support in the Posgrado en Astrof\'{i}sica graduate program at Instituto de Astronom\'{i}a, UNAM. JH acknowledges support from the National Research Council of M\'exico (CONACyT) project No. 86372 and the PAPIIT UNAM projects  IA102921 and IA102319. EMM acknowledges the receipt of the grant from the Abdus Salam International Centre for Theoretical Physics, Trieste, Italy, and also the receipt of a postdoctoral grant from CONACYT. ARL acknowledges financial support provided in Chile by Agencia Nacional de Investigaci\'on y Desarrollo (ANID) through the FONDECYT project 1170476. C.R-Z. acknowledges support from project CONACYT CB2018A1-S-9754, and PAPIIT UNAM IN112620. KPR acknowledges the support of the ANID FONDECYT Iniciaci\'on 11201161 grant. MT acknowledges support from PAPIIT UNAM project No. IN107519. An anonymous referee provided comments and suggestions that improved the content and presentation of this work.\\
This paper includes data collected with the TESS mission, obtained from the MAST data archive at the Space Telescope Science Institute (STScI). Funding for the TESS mission is provided by the NASA Explorer Program.\\
Funding for the Sloan Digital Sky Survey IV has been provided by the Alfred P. Sloan Foundation, the U.S. Department of Energy Office of Science, and the Participating Institutions. SDSS-IV acknowledges support and resources from the Center for High Performance Computing  at the University of Utah. The SDSS 
website is www.sdss.org. SDSS-IV is managed by the Astrophysical Research Consortium for the Participating Institutions of the SDSS Collaboration including the Brazilian Participation Group, the Carnegie Institution for Science, Carnegie Mellon University, Center for 
Astrophysics | Harvard \& Smithsonian, the Chilean Participation 
Group, the French Participation Group, Instituto de Astrof\'isica de Canarias, The Johns Hopkins University, Kavli Institute for the Physics and Mathematics of the Universe (IPMU) / University of 
Tokyo, the Korean Participation Group, Lawrence Berkeley National Laboratory, Leibniz Institut f\"ur Astrophysik Potsdam (AIP),  Max-Planck-Institut f\"ur Astronomie (MPIA Heidelberg), 
Max-Planck-Institut f\"ur Astrophysik (MPA Garching), 
Max-Planck-Institut f\"ur Extraterrestrische Physik (MPE), 
National Astronomical Observatories of China, New Mexico State University, New York University, University of Notre Dame, Observat\'ario Nacional / MCTI, The Ohio State University, Pennsylvania State University, Shanghai Astronomical Observatory, United Kingdom Participation Group, Universidad Nacional Aut\'onoma de M\'exico, University of Arizona, University of Colorado Boulder, University of Oxford, University of 
Portsmouth, University of Utah, University of Virginia, University of Washington, University of Wisconsin, Vanderbilt University, 
and Yale University.

\vspace{5mm}

\facilities{TESS, ASAS-SN}

\software{Numpy \citep{Harris2020}, Matplotlib \citep{Hunter2007}, Astropy \citep{2013A&A...558A..33A}, Astrocut\citep{TESSCut}, Photutils \citep{Bradley_2019}, lmfit \citep{Newville_2014}}

\bibliography{temp}{}

\begin{thebibliography}{}
\expandafter\ifx\csname natexlab\endcsname\relax\def\natexlab#1{#1}\fi
\providecommand{\url}[1]{\href{#1}{#1}}
\providecommand{\dodoi}[1]{doi:~\href{http://doi.org/#1}{\nolinkurl{#1}}}
\providecommand{\doeprint}[1]{\href{http://ascl.net/#1}{\nolinkurl{http://ascl.net/#1}}}
\providecommand{\doarXiv}[1]{\href{https://arxiv.org/abs/#1}{\nolinkurl{https://arxiv.org/abs/#1}}}

\bibitem[{{Amard} {et~al.}(2016){Amard}, {Palacios}, {Charbonnel}, {Gallet}, \&
  {Bouvier}}]{Amard2016}
{Amard}, L., {Palacios}, A., {Charbonnel}, C., {Gallet}, F., \& {Bouvier}, J.
  2016, \aap, 587, A105, \dodoi{10.1051/0004-6361/201527349}

\bibitem[{{Amard} {et~al.}(2019){Amard}, {Palacios}, {Charbonnel}, {Gallet},
  {Georgy}, {Lagarde}, \& {Siess}}]{Amard2019}
{Amard}, L., {Palacios}, A., {Charbonnel}, C., {et~al.} 2019, \aap, 631, A77,
  \dodoi{10.1051/0004-6361/201935160}

\bibitem[{{Anderson}(1976)}]{Anderson1976}
{Anderson}, G.~M. 1976, \gca, 40, 1533, \dodoi{10.1016/0016-7037(76)90092-2}

\bibitem[{{Artemenko} {et~al.}(2012){Artemenko}, {Grankin}, \&
  {Petrov}}]{Artemenko2012}
{Artemenko}, S.~A., {Grankin}, K.~N., \& {Petrov}, P.~P. 2012, Astronomy
  Letters, 38, 783, \dodoi{10.1134/S1063773712110011}

\bibitem[{{Astropy Collaboration} {et~al.}(2013){Astropy Collaboration},
  {Robitaille}, {Tollerud}, {Greenfield}, {Droettboom}, {Bray}, {Aldcroft},
  {Davis}, {Ginsburg}, {Price-Whelan}, {Kerzendorf}, {Conley}, {Crighton},
  {Barbary}, {Muna}, {Ferguson}, {Grollier}, {Parikh}, {Nair}, {Unther},
  {Deil}, {Woillez}, {Conseil}, {Kramer}, {Turner}, {Singer}, {Fox}, {Weaver},
  {Zabalza}, {Edwards}, {Azalee Bostroem}, {Burke}, {Casey}, {Crawford},
  {Dencheva}, {Ely}, {Jenness}, {Labrie}, {Lim}, {Pierfederici}, {Pontzen},
  {Ptak}, {Refsdal}, {Servillat}, \& {Streicher}}]{2013A&A...558A..33A}
{Astropy Collaboration}, {Robitaille}, T.~P., {Tollerud}, E.~J., {et~al.} 2013,
  \aap, 558, A33, \dodoi{10.1051/0004-6361/201322068}

\bibitem[{Baluev(2008)}]{Baluev_2008}
Baluev, R.~V. 2008, Monthly Notices of the Royal Astronomical Society, 385,
  1279, \dodoi{10.1111/j.1365-2966.2008.12689.x}

\bibitem[{{Baraffe} {et~al.}(1998){Baraffe}, {Chabrier}, {Allard}, \&
  {Hauschildt}}]{Baraffe1998}
{Baraffe}, I., {Chabrier}, G., {Allard}, F., \& {Hauschildt}, P.~H. 1998, \aap,
  337, 403.
\newblock \doarXiv{astro-ph/9805009}

\bibitem[{Baraffe {et~al.}(2015)Baraffe, Homeier, Allard, \&
  Chabrier}]{Baraffe_2015}
Baraffe, I., Homeier, D., Allard, F., \& Chabrier, G. 2015, A\&A, 577, A42,
  \dodoi{10.1051/0004-6361/201425481}

\bibitem[{Barnes {et~al.}(2005)Barnes, Cameron, Donati, James, Marsden, \&
  Petit}]{Barnes2005}
Barnes, J.~R., Cameron, A.~C., Donati, J.-F., {et~al.} 2005, Monthly Notices of
  the Royal Astronomical Society: Letters, 357, L1,
  \dodoi{10.1111/j.1745-3933.2005.08587.x}

\bibitem[{{Biazzo} {et~al.}(2009){Biazzo}, {Melo}, {Pasquini}, {Randich},
  {Bouvier}, \& {Delfosse}}]{Biazzo_2009}
{Biazzo}, K., {Melo}, C. H.~F., {Pasquini}, L., {et~al.} 2009, A\&A, 508, 1301,
  \dodoi{10.1051/0004-6361/200913125}

\bibitem[{Blanc {et~al.}(2011)Blanc, Covey, \& Stassun}]{Le_Blanc_2011}
Blanc, T. S.~L., Covey, K.~R., \& Stassun, K.~G. 2011, The Astronomical
  Journal, 142, 55, \dodoi{10.1088/0004-6256/142/2/55}

\bibitem[{{Bouvier}(2013)}]{Bouvier2013}
{Bouvier}, J. 2013, EAS Publications Series, 62, 143,
  \dodoi{10.1051/eas/1362005}

\bibitem[{{Bouvier} {et~al.}(1993){Bouvier}, {Cabrit}, {Fernandez}, {Martin},
  \& {Matthews}}]{Bouvier1993}
{Bouvier}, J., {Cabrit}, S., {Fernandez}, M., {Martin}, E.~L., \& {Matthews},
  J.~M. 1993, \aap, 272, 176

\bibitem[{{Bouvier} {et~al.}(2014){Bouvier}, {Matt}, {Mohanty}, {Scholz},
  {Stassun}, \& {Zanni}}]{Bouvier2014}
{Bouvier}, J., {Matt}, S.~P., {Mohanty}, S., {et~al.} 2014, in Protostars and
  Planets VI, ed. H.~{Beuther}, R.~S. {Klessen}, C.~P. {Dullemond}, \&
  T.~{Henning}, 433, \dodoi{10.2458/azu_uapress_9780816531240-ch019}

\bibitem[{{Bouvier} {et~al.}(2016){Bouvier}, {Lanzafame}, {Venuti}, {Klutsch},
  {Jeffries}, {Frasca}, {Moraux}, {Biazzo}, {Messina}, {Micela}, {Randich},
  {Stauffer}, {Cody}, {Flaccomio}, {Gilmore}, {Bayo}, {Bensby}, {Bragaglia},
  {Carraro}, {Casey}, {Costado}, {Damiani}, {Delgado Mena}, {Donati},
  {Franciosini}, {Hourihane}, {Koposov}, {Lardo}, {Lewis}, {Magrini}, {Monaco},
  {Morbidelli}, {Prisinzano}, {Sacco}, {Sbordone}, {Sousa}, {Vallenari},
  {Worley}, {Zaggia}, \& {Zwitter}}]{Bouvier2016}
{Bouvier}, J., {Lanzafame}, A.~C., {Venuti}, L., {et~al.} 2016, \aap, 590, A78,
  \dodoi{10.1051/0004-6361/201628336}

\bibitem[{Bradley {et~al.}(2019)Bradley, Sip{\H o}cz, Robitaille, Tollerud,
  Vin{\'{\i}}cius, Deil, Barbary, G{\"u}nther, Cara, Busko, Conseil,
  Droettboom, Bostroem, Bray, Bratholm, Wilson, Craig, Barentsen, Pascual,
  Donath, Greco, Perren, Lim, \& Kerzendorf}]{Bradley_2019}
Bradley, L., Sip{\H o}cz, B., Robitaille, T., {et~al.} 2019, astropy/photutils:
  v0.6, \dodoi{10.5281/zenodo.2533376}

\bibitem[{{Brasseur} {et~al.}(2019){Brasseur}, {Phillip}, {Fleming},
  {Mullally}, \& {White}}]{TESSCut}
{Brasseur}, C.~E., {Phillip}, C., {Fleming}, S.~W., {Mullally}, S.~E., \&
  {White}, R.~L. 2019, {Astrocut: Tools for creating cutouts of TESS images}.
\newblock \doeprint{1905.007}

\bibitem[{Brice{\~n}o(2008)}]{Briceno2008}
Brice{\~n}o, C. 2008, Handb. Star Form. Reg. Vol. I North. Sky ASP Monogr.
  Publ. Vol. 4. Ed. by Bo Reipurth, p.838, 4, 838.
\newblock \doarXiv{0810.2294}

\bibitem[{Brice{\~{n}}o {et~al.}(2019)Brice{\~{n}}o, Calvet, Hern{\'{a}}ndez,
  Vivas, Mateu, Downes, Loerincs, P{\'{e}}rez-Blanco, Berlind, Espaillat,
  Allen, Hartmann, Mateo, \& III}]{Briceno_2019}
Brice{\~{n}}o, C., Calvet, N., Hern{\'{a}}ndez, J., {et~al.} 2019, The
  Astronomical Journal, 157, 85, \dodoi{10.3847/1538-3881/aaf79b}

\bibitem[{{Brown} {et~al.}(2016){Brown}, {Breeveld}, {Roming}, \&
  {Siegel}}]{Brown2016}
{Brown}, P.~J., {Breeveld}, A., {Roming}, P. W.~A., \& {Siegel}, M. 2016, \aj,
  152, 102, \dodoi{10.3847/0004-6256/152/4/102}

\bibitem[{Carpenter {et~al.}(2001)Carpenter, Hillenbrand, \&
  Skrutskie}]{Carpenter2001}
Carpenter, J.~M., Hillenbrand, L.~A., \& Skrutskie, M.~F. 2001, Astron. J.,
  121, 3160, \dodoi{10.1086/321086}

\bibitem[{Carpenter {et~al.}(2006)Carpenter, Mamajek, Hillenbrand, \&
  Meyer}]{Carpenter_2006}
Carpenter, J.~M., Mamajek, E.~E., Hillenbrand, L.~A., \& Meyer, M.~R. 2006, The
  Astrophysical Journal, 651, L49, \dodoi{10.1086/509121}

\bibitem[{Carroll(1933)}]{Carroll1933}
Carroll, J.~A. 1933, Mon. Not. R. Astron. Soc., 93, 478,
  \dodoi{10.1093/mnras/93.7.478}

\bibitem[{Cody \& Hillenbrand(2010)}]{Cody2010}
Cody, A.~M., \& Hillenbrand, L.~A. 2010, Astrophys. J. Suppl. Ser., 191, 389,
  \dodoi{10.1088/0067-0049/191/2/389}

\bibitem[{{Cody} \& {Hillenbrand}(2018)}]{Cody2018}
{Cody}, A.~M., \& {Hillenbrand}, L.~A. 2018, \aj, 156, 71,
  \dodoi{10.3847/1538-3881/aacead}

\bibitem[{{Cody} {et~al.}(2014){Cody}, {Stauffer}, {Baglin}, {Micela},
  {Rebull}, {Flaccomio}, {Morales-Calder{\'o}n}, {Aigrain}, {Bouvier},
  {Hillenbrand}, {Gutermuth}, {Song}, {Turner}, {Alencar}, {Zwintz},
  {Plavchan}, {Carpenter}, {Findeisen}, {Carey}, {Terebey}, {Hartmann},
  {Calvet}, {Teixeira}, {Vrba}, {Wolk}, {Covey}, {Poppenhaeger}, {G{\"u}nther},
  {Forbrich}, {Whitney}, {Affer}, {Herbst}, {Hora}, {Barrado}, {Holtzman},
  {Marchis}, {Wood}, {Medeiros Guimar{\~a}es}, {Lillo Box}, {Gillen},
  {McQuillan}, {Espaillat}, {Allen}, {D'Alessio}, \& {Favata}}]{Cody2014}
{Cody}, A.~M., {Stauffer}, J., {Baglin}, A., {et~al.} 2014, \aj, 147, 82,
  \dodoi{10.1088/0004-6256/147/4/82}

\bibitem[{Cottaar {et~al.}(2014)Cottaar, Covey, Meyer, Nidever, Stassun,
  Foster, Tan, Chojnowski, da~Rio, Flaherty, Frinchaboy, Skrutskie, Majewski,
  Wilson, \& Zasowski}]{Cottaar2014}
Cottaar, M., Covey, K.~R., Meyer, M.~R., {et~al.} 2014, The Astrophysical
  Journal, 794, 125, \dodoi{10.1088/0004-637x/794/2/125}

\bibitem[{Cottle {et~al.}(2018)Cottle, Covey, Su{\'{a}}rez,
  Rom{\'{a}}n-Z{\'{u}}{\~{n}}iga, Schlafly, Downes, Ybarra, Hernandez, Stassun,
  Stringfellow, Getman, Feigelson, Borissova, Kim, Roman-Lopes, {Da Rio}, {De
  Lee}, Frinchaboy, Kounkel, Majewski, Mennickent, Nidever, Nitschelm, Pan,
  Shetrone, Zasowski, Chambers, Magnier, \& Valenti}]{Cottle2018}
Cottle, J., Covey, K.~R., Su{\'{a}}rez, G., {et~al.} 2018, Astrophys. J. Suppl.
  Ser., 236, 27, \dodoi{10.3847/1538-4365/aabada}

\bibitem[{{Cutri} {et~al.}(2003){Cutri}, {Skrutskie}, {van Dyk}, {Beichman},
  {Carpenter}, {Chester}, {Cambresy}, {Evans}, {Fowler}, {Gizis}, {Howard},
  {Huchra}, {Jarrett}, {Kopan}, {Kirkpatrick}, {Light}, {Marsh}, {McCallon},
  {Schneider}, {Stiening}, {Sykes}, {Weinberg}, {Wheaton}, {Wheelock}, \&
  {Zacarias}}]{2MASS2003}
{Cutri}, R.~M., {Skrutskie}, M.~F., {van Dyk}, S., {et~al.} 2003, VizieR Online
  Data Catalog, II/246

\bibitem[{Davies {et~al.}(2014)Davies, Gregory, \& Greaves}]{Davies2014}
Davies, C.~L., Gregory, S.~G., \& Greaves, J.~S. 2014, Mon. Not. R. Astron.
  Soc., 444, 1157, \dodoi{10.1093/mnras/stu1488}

\bibitem[{{Dotter}(2016)}]{Dotter2016}
{Dotter}, A. 2016, \apjs, 222, 8, \dodoi{10.3847/0067-0049/222/1/8}

\bibitem[{{Fedele} {et~al.}(2010){Fedele}, {van den Ancker}, {Henning},
  {Jayawardhana}, \& {Oliveira}}]{Fedele2010}
{Fedele}, D., {van den Ancker}, M.~E., {Henning}, T., {Jayawardhana}, R., \&
  {Oliveira}, J.~M. 2010, \aap, 510, A72, \dodoi{10.1051/0004-6361/200912810}

\bibitem[{{Fitzpatrick} {et~al.}(2019){Fitzpatrick}, {Massa}, {Gordon},
  {Bohlin}, \& {Clayton}}]{Fitzpatrick2019}
{Fitzpatrick}, E.~L., {Massa}, D., {Gordon}, K.~D., {Bohlin}, R., \& {Clayton},
  G.~C. 2019, \apj, 886, 108, \dodoi{10.3847/1538-4357/ab4c3a}

\bibitem[{{Frasca} {et~al.}(2009){Frasca}, {Covino}, {Spezzi}, {Alcal\'a},
  {Marilli}, {F"ur\'esz}, \& {Gandolfi}}]{Frasca_2009}
{Frasca}, A., {Covino}, E., {Spezzi}, L., {et~al.} 2009, A\&A, 508, 1313,
  \dodoi{10.1051/0004-6361/200913327}

\bibitem[{{Gaia Collaboration} {et~al.}(2020){Gaia Collaboration}, {Brown},
  {Vallenari}, {Prusti}, {de Bruijne}, {Babusiaux}, \& {Biermann}}]{GAIA_EDR3}
{Gaia Collaboration}, {Brown}, A.~G.~A., {Vallenari}, A., {et~al.} 2020, arXiv
  e-prints, arXiv:2012.01533.
\newblock \doarXiv{2012.01533}

\bibitem[{{Gaia Collaboration} {et~al.}(2018){Gaia Collaboration}, {Brown, A.
  G. A.}, {Vallenari, A.}, {Prusti, T.}, {de Bruijne, J. H. J.}, {Babusiaux,
  C.}, {Bailer-Jones, C. A. L.}, {Biermann, M.}, {Evans, D. W.}, {Eyer, L.},
  {Jansen, F.}, {Jordi, C.}, {Klioner, S. A.}, {Lammers, U.}, {Lindegren, L.},
  {Luri, X.}, {Mignard, F.}, {Panem, C.}, {Pourbaix, D.}, {Randich, S.},
  {Sartoretti, P.}, {Siddiqui, H. I.}, {Soubiran, C.}, {van Leeuwen, F.},
  {Walton, N. A.}, {Arenou, F.}, {Bastian, U.}, {Cropper, M.}, {Drimmel, R.},
  {Katz, D.}, {Lattanzi, M. G.}, {Bakker, J.}, {Cacciari, C.}, {Casta\~neda,
  J.}, {Chaoul, L.}, {Cheek, N.}, {De Angeli, F.}, {Fabricius, C.}, {Guerra,
  R.}, {Holl, B.}, {Masana, E.}, {Messineo, R.}, {Mowlavi, N.}, {Nienartowicz,
  K.}, {Panuzzo, P.}, {Portell, J.}, {Riello, M.}, {Seabroke, G. M.}, {Tanga,
  P.}, {Th\'evenin, F.}, {Gracia-Abril, G.}, {Comoretto, G.},
  {Garcia-Reinaldos, M.}, {Teyssier, D.}, {Altmann, M.}, {Andrae, R.}, {Audard,
  M.}, {Bellas-Velidis, I.}, {Benson, K.}, {Berthier, J.}, {Blomme, R.},
  {Burgess, P.}, {Busso, G.}, {Carry, B.}, {Cellino, A.}, {Clementini, G.},
  {Clotet, M.}, {Creevey, O.}, {Davidson, M.}, {De Ridder, J.}, {Delchambre,
  L.}, {Dell\'{}Oro, A.}, {Ducourant, C.}, {Fern\'andez-Hern\'andez, J.},
  {Fouesneau, M.}, {Fr\'emat, Y.}, {Galluccio, L.}, {Garc\'{\i}a-Torres, M.},
  {Gonz\'alez-N\'u\~nez, J.}, {Gonz\'alez-Vidal, J. J.}, {Gosset, E.}, {Guy, L.
  P.}, {Halbwachs, J.-L.}, {Hambly, N. C.}, {Harrison, D. L.}, {Hern\'andez,
  J.}, {Hestroffer, D.}, {Hodgkin, S. T.}, {Hutton, A.}, {Jasniewicz, G.},
  {Jean-Antoine-Piccolo, A.}, {Jordan, S.}, {Korn, A. J.}, {Krone-Martins, A.},
  {Lanzafame, A. C.}, {Lebzelter, T.}, {L\"offler, W.}, {Manteiga, M.},
  {Marrese, P. M.}, {Mart\'{\i}n-Fleitas, J. M.}, {Moitinho, A.}, {Mora, A.},
  {Muinonen, K.}, {Osinde, J.}, {Pancino, E.}, {Pauwels, T.}, {Petit, J.-M.},
  {Recio-Blanco, A.}, {Richards, P. J.}, {Rimoldini, L.}, {Robin, A. C.},
  {Sarro, L. M.}, {Siopis, C.}, {Smith, M.}, {Sozzetti, A.}, {S\"uveges, M.},
  {Torra, J.}, {van Reeven, W.}, {Abbas, U.}, {Abreu Aramburu, A.}, {Accart,
  S.}, {Aerts, C.}, {Altavilla, G.}, {\'Alvarez, M. A.}, {Alvarez, R.}, {Alves,
  J.}, {Anderson, R. I.}, {Andrei, A. H.}, {Anglada Varela, E.}, {Antiche, E.},
  {Antoja, T.}, {Arcay, B.}, {Astraatmadja, T. L.}, {Bach, N.}, {Baker, S. G.},
  {Balaguer-N\'u\~nez, L.}, {Balm, P.}, {Barache, C.}, {Barata, C.}, {Barbato,
  D.}, {Barblan, F.}, {Barklem, P. S.}, {Barrado, D.}, {Barros, M.}, {Barstow,
  M. A.}, {Bartholom\'e Mu\~noz, S.}, {Bassilana, J.-L.}, {Becciani, U.},
  {Bellazzini, M.}, {Berihuete, A.}, {Bertone, S.}, {Bianchi, L.}, {Bienaym\'e,
  O.}, {Blanco-Cuaresma, S.}, {Boch, T.}, {Boeche, C.}, {Bombrun, A.},
  {Borrachero, R.}, {Bossini, D.}, {Bouquillon, S.}, {Bourda, G.}, {Bragaglia,
  A.}, {Bramante, L.}, {Breddels, M. A.}, {Bressan, A.}, {Brouillet, N.},
  {Br\"usemeister, T.}, {Brugaletta, E.}, {Bucciarelli, B.}, {Burlacu, A.},
  {Busonero, D.}, {Butkevich, A. G.}, {Buzzi, R.}, {Caffau, E.}, {Cancelliere,
  R.}, {Cannizzaro, G.}, {Cantat-Gaudin, T.}, {Carballo, R.}, {Carlucci, T.},
  {Carrasco, J. M.}, {Casamiquela, L.}, {Castellani, M.}, {Castro-Ginard, A.},
  {Charlot, P.}, {Chemin, L.}, {Chiavassa, A.}, {Cocozza, G.}, {Costigan, G.},
  {Cowell, S.}, {Crifo, F.}, {Crosta, M.}, {Crowley, C.}, {Cuypers+, J.},
  {Dafonte, C.}, {Damerdji, Y.}, {Dapergolas, A.}, {David, P.}, {David, M.},
  {de Laverny, P.}, {De Luise, F.}, {De March, R.}, {de Martino, D.}, {de
  Souza, R.}, {de Torres, A.}, {Debosscher, J.}, {del Pozo, E.}, {Delbo, M.},
  {Delgado, A.}, {Delgado, H. E.}, {Di Matteo, P.}, {Diakite, S.}, {Diener,
  C.}, {Distefano, E.}, {Dolding, C.}, {Drazinos, P.}, {Dur\'an, J.},
  {Edvardsson, B.}, {Enke, H.}, {Eriksson, K.}, {Esquej, P.}, {Eynard Bontemps,
  G.}, {Fabre, C.}, {Fabrizio, M.}, {Faigler, S.}, {Falc\~ao, A. J.}, {Farr\`as
  Casas, M.}, {Federici, L.}, {Fedorets, G.}, {Fernique, P.}, {Figueras, F.},
  {Filippi, F.}, {Findeisen, K.}, {Fonti, A.}, {Fraile, E.}, {Fraser, M.},
  {Fr\'ezouls, B.}, {Gai, M.}, {Galleti, S.}, {Garabato, D.},
  {Garc\'{\i}a-Sedano, F.}, {Garofalo, A.}, {Garralda, N.}, {Gavel, A.},
  {Gavras, P.}, {Gerssen, J.}, {Geyer, R.}, {Giacobbe, P.}, {Gilmore, G.},
  {Girona, S.}, {Giuffrida, G.}, {Glass, F.}, {Gomes, M.}, {Granvik, M.},
  {Gueguen, A.}, {Guerrier, A.}, {Guiraud, J.}, {Guti\'errez-S\'anchez, R.},
  {Haigron, R.}, {Hatzidimitriou, D.}, {Hauser, M.}, {Haywood, M.}, {Heiter,
  U.}, {Helmi, A.}, {Heu, J.}, {Hilger, T.}, {Hobbs, D.}, {Hofmann, W.},
  {Holland, G.}, {Huckle, H. E.}, {Hypki, A.}, {Icardi, V.}, {Jan\ss{}en, K.},
  {Jevardat de Fombelle, G.}, {Jonker, P. G.}, {Juh\'asz, \'A. L.}, {Julbe,
  F.}, {Karampelas, A.}, {Kewley, A.}, {Klar, J.}, {Kochoska, A.}, {Kohley,
  R.}, {Kolenberg, K.}, {Kontizas, M.}, {Kontizas, E.}, {Koposov, S. E.},
  {Kordopatis, G.}, {Kostrzewa-Rutkowska, Z.}, {Koubsky, P.}, {Lambert, S.},
  {Lanza, A. F.}, {Lasne, Y.}, {Lavigne, J.-B.}, {Le Fustec, Y.}, {Le
  Poncin-Lafitte, C.}, {Lebreton, Y.}, {Leccia, S.}, {Leclerc, N.},
  {Lecoeur-Taibi, I.}, {Lenhardt, H.}, {Leroux, F.}, {Liao, S.}, {Licata, E.},
  {Lindstr\o{}m, H. E. P.}, {Lister, T. A.}, {Livanou, E.}, {Lobel, A.},
  {L\'opez, M.}, {Managau, S.}, {Mann, R. G.}, {Mantelet, G.}, {Marchal, O.},
  {Marchant, J. M.}, {Marconi, M.}, {Marinoni, S.}, {Marschalk\'o, G.},
  {Marshall, D. J.}, {Martino, M.}, {Marton, G.}, {Mary, N.}, {Massari, D.},
  {Matijevic, G.}, {Mazeh, T.}, {McMillan, P. J.}, {Messina, S.}, {Michalik,
  D.}, {Millar, N. R.}, {Molina, D.}, {Molinaro, R.}, {Moln\'ar, L.},
  {Montegriffo, P.}, {Mor, R.}, {Morbidelli, R.}, {Morel, T.}, {Morris, D.},
  {Mulone, A. F.}, {Muraveva, T.}, {Musella, I.}, {Nelemans, G.}, {Nicastro,
  L.}, {Noval, L.}, {O\'{}Mullane, W.}, {Ord\'enovic, C.}, {Ord\'o\~nez-Blanco,
  D.}, {Osborne, P.}, {Pagani, C.}, {Pagano, I.}, {Pailler, F.}, {Palacin, H.},
  {Palaversa, L.}, {Panahi, A.}, {Pawlak, M.}, {Piersimoni, A. M.}, {Pineau,
  F.-X.}, {Plachy, E.}, {Plum, G.}, {Poggio, E.}, {Poujoulet, E.}, {Prsa, A.},
  {Pulone, L.}, {Racero, E.}, {Ragaini, S.}, {Rambaux, N.}, {Ramos-Lerate, M.},
  {Regibo, S.}, {Reyl\'e, C.}, {Riclet, F.}, {Ripepi, V.}, {Riva, A.}, {Rivard,
  A.}, {Rixon, G.}, {Roegiers, T.}, {Roelens, M.}, {Romero-G\'omez, M.},
  {Rowell, N.}, {Royer, F.}, {Ruiz-Dern, L.}, {Sadowski, G.}, {Sagrist\`a
  Sell\'es, T.}, {Sahlmann, J.}, {Salgado, J.}, {Salguero, E.}, {Sanna, N.},
  {Santana-Ros, T.}, {Sarasso, M.}, {Savietto, H.}, {Schultheis, M.}, {Sciacca,
  E.}, {Segol, M.}, {Segovia, J. C.}, {S\'egransan, D.}, {Shih, I-C.},
  {Siltala, L.}, {Silva, A. F.}, {Smart, R. L.}, {Smith, K. W.}, {Solano, E.},
  {Solitro, F.}, {Sordo, R.}, {Soria Nieto, S.}, {Souchay, J.}, {Spagna, A.},
  {Spoto, F.}, {Stampa, U.}, {Steele, I. A.}, {Steidelm\"uller, H.},
  {Stephenson, C. A.}, {Stoev, H.}, {Suess, F. F.}, {Surdej, J.}, {Szabados,
  L.}, {Szegedi-Elek, E.}, {Tapiador, D.}, {Taris, F.}, {Tauran, G.}, {Taylor,
  M. B.}, {Teixeira, R.}, {Terrett, D.}, {Teyssandier, P.}, {Thuillot, W.},
  {Titarenko, A.}, {Torra Clotet, F.}, {Turon, C.}, {Ulla, A.}, {Utrilla, E.},
  {Uzzi, S.}, {Vaillant, M.}, {Valentini, G.}, {Valette, V.}, {van Elteren,
  A.}, {Van Hemelryck, E.}, {van Leeuwen, M.}, {Vaschetto, M.}, {Vecchiato,
  A.}, {Veljanoski, J.}, {Viala, Y.}, {Vicente, D.}, {Vogt, S.}, {von Essen,
  C.}, {Voss, H.}, {Votruba, V.}, {Voutsinas, S.}, {Walmsley, G.}, {Weiler,
  M.}, {Wertz, O.}, {Wevers, T.}, {Wyrzykowski, L.}, {Yoldas, A.}, {Zerjal,
  M.}, {Ziaeepour, H.}, {Zorec, J.}, {Zschocke, S.}, {Zucker, S.}, {Zurbach,
  C.}, \& {Zwitter, T.}}]{Gaia_2018}
{Gaia Collaboration}, {Brown, A. G. A.}, {Vallenari, A.}, {et~al.} 2018, A\&A,
  616, A1, \dodoi{10.1051/0004-6361/201833051}

\bibitem[{Gallet \& Bouvier(2013)}]{Gallet2013}
Gallet, F., \& Bouvier, J. 2013, Astron. Astrophys., 556, A36,
  \dodoi{10.1051/0004-6361/201321302}

\bibitem[{Gallet \& Bouvier(2015)}]{Gallet_2015}
---. 2015, A\&A, 577, A98, \dodoi{10.1051/0004-6361/201525660}

\bibitem[{{Gallet} {et~al.}(2019){Gallet}, {Zanni}, \& {Amard}}]{Gallet2019}
{Gallet}, F., {Zanni}, C., \& {Amard}, L. 2019, \aap, 632, A6,
  \dodoi{10.1051/0004-6361/201935432}

\bibitem[{{Garc{\'\i}a P{\'e}rez} {et~al.}(2016){Garc{\'\i}a P{\'e}rez},
  {Allende Prieto}, {Holtzman}, {Shetrone}, {M{\'e}sz{\'a}ros}, {Bizyaev},
  {Carrera}, {Cunha}, {Garc{\'\i}a-Hern{\'a}ndez}, {Johnson}, {Majewski},
  {Nidever}, {Schiavon}, {Shane}, {Smith}, {Sobeck}, {Troup}, {Zamora},
  {Weinberg}, {Bovy}, {Eisenstein}, {Feuillet}, {Frinchaboy}, {Hayden},
  {Hearty}, {Nguyen}, {O'Connell}, {Pinsonneault}, {Wilson}, \&
  {Zasowski}}]{Garcia-Perez_2016}
{Garc{\'\i}a P{\'e}rez}, A.~E., {Allende Prieto}, C., {Holtzman}, J.~A.,
  {et~al.} 2016, \aj, 151, 144, \dodoi{10.3847/0004-6256/151/6/144}

\bibitem[{{Gilhool} {et~al.}(2018){Gilhool}, {Blake}, {Terrien}, {Bender},
  {Mahadevan}, \& {Deshpande}}]{Gilhool2018}
{Gilhool}, S.~H., {Blake}, C.~H., {Terrien}, R.~C., {et~al.} 2018, \aj, 155,
  38, \dodoi{10.3847/1538-3881/aa9c7c}

\bibitem[{{Godoy-Rivera} {et~al.}(2021){Godoy-Rivera}, {Pinsonneault}, \&
  {Rebull}}]{GodoyRivera2021}
{Godoy-Rivera}, D., {Pinsonneault}, M.~H., \& {Rebull}, L.~M. 2021, arXiv
  e-prints, arXiv:2101.01183.
\newblock \doarXiv{2101.01183}

\bibitem[{{Gro{\ss}schedl} {et~al.}(2018){Gro{\ss}schedl}, {Alves}, {Meingast},
  {Ackerl}, {Ascenso}, {Bouy}, {Burkert}, {Forbrich}, {F{\"u}rnkranz},
  {Goodman}, {Hacar}, {Herbst-Kiss}, {Lada}, {Larreina}, {Leschinski},
  {Lombardi}, {Moitinho}, {Mortimer}, \& {Zari}}]{Grossschedl2018}
{Gro{\ss}schedl}, J.~E., {Alves}, J., {Meingast}, S., {et~al.} 2018, \aap, 619,
  A106, \dodoi{10.1051/0004-6361/201833901}

\bibitem[{Harris {et~al.}(2020)Harris, Millman, van~der Walt, Gommers,
  Virtanen, Cournapeau, Wieser, Taylor, Berg, Smith, Kern, Picus, Hoyer, van
  Kerkwijk, Brett, Haldane, del R{'{\i}}o, Wiebe, Peterson,
  G{'{e}}rard-Marchant, Sheppard, Reddy, Weckesser, Abbasi, Gohlke, \&
  Oliphant}]{Harris2020}
Harris, C.~R., Millman, K.~J., van~der Walt, S.~J., {et~al.} 2020, Nature, 585,
  357, \dodoi{10.1038/s41586-020-2649-2}

\bibitem[{{Hartman} {et~al.}(2010){Hartman}, {Bakos}, {Kov{\'a}cs}, \&
  {Noyes}}]{Hartman_2010}
{Hartman}, J.~D., {Bakos}, G.~{\'A}., {Kov{\'a}cs}, G., \& {Noyes}, R.~W. 2010,
  \mnras, 408, 475, \dodoi{10.1111/j.1365-2966.2010.17147.x}

\bibitem[{{Herbst} {et~al.}(2002){Herbst}, {Bailer-Jones}, {Mundt},
  {Meisenheimer}, \& {Wackermann}}]{Herbst_2002}
{Herbst}, W., {Bailer-Jones}, C. A.~L., {Mundt}, R., {Meisenheimer}, K., \&
  {Wackermann}, R. 2002, A\&A, 396, 513, \dodoi{10.1051/0004-6361:20021362}

\bibitem[{{Hernandez} {et~al.}(2017){Hernandez}, {Brice{\~n}o}, {Calvet},
  {Hartmann}, {Berlind}, \& {Luhman}}]{Hernandez2017}
{Hernandez}, J., {Brice{\~n}o}, C., {Calvet}, N., {et~al.} 2017, {SPTCLASS:
  SPecTral CLASSificator code}.
\newblock \doeprint{1705.005}

\bibitem[{{Hern{\'a}ndez} {et~al.}(2004){Hern{\'a}ndez}, {Calvet},
  {Brice{\~n}o}, {Hartmann}, \& {Berlind}}]{Hernandez2004}
{Hern{\'a}ndez}, J., {Calvet}, N., {Brice{\~n}o}, C., {Hartmann}, L., \&
  {Berlind}, P. 2004, \aj, 127, 1682, \dodoi{10.1086/381908}

\bibitem[{Hernandez {et~al.}(2007)Hernandez, Hartmann, Megeath, Gutermuth,
  Muzerolle, Calvet, Vivas, Briceno, Allen, Stauffer, Young, \&
  Fazio}]{Hernandez2007}
Hernandez, J., Hartmann, L., Megeath, T., {et~al.} 2007, The Astrophysical
  Journal, 662, 1067, \dodoi{10.1086/513735}

\bibitem[{Hern{\'{a}}ndez {et~al.}(2014)Hern{\'{a}}ndez, Calvet, Perez,
  Brice{\~{n}}o, Olguin, Contreras, Hartmann, Allen, Espaillat, \&
  Hernan}]{Hernandez_2014}
Hern{\'{a}}ndez, J., Calvet, N., Perez, A., {et~al.} 2014, The Astrophysical
  Journal, 794, 36, \dodoi{10.1088/0004-637x/794/1/36}

\bibitem[{Hunter(2007)}]{Hunter2007}
Hunter, J.~D. 2007, Computing in Science \& Engineering, 9, 90,
  \dodoi{10.1109/MCSE.2007.55}

\bibitem[{{Husser} {et~al.}(2013){Husser}, {Wende-von Berg}, {Dreizler},
  {Homeier}, {Reiners}, {Barman}, \& {Hauschildt}}]{Husser2013}
{Husser}, T.~O., {Wende-von Berg}, S., {Dreizler}, S., {et~al.} 2013, \aap,
  553, A6, \dodoi{10.1051/0004-6361/201219058}

\bibitem[{{Jackson} \& {Jeffries}(2014)}]{Jackson2014}
{Jackson}, R.~J., \& {Jeffries}, R.~D. 2014, \mnras, 441, 2111,
  \dodoi{10.1093/mnras/stu651}

\bibitem[{{Jackson, R. J.} {et~al.}(2016){Jackson, R. J.}, {Jeffries, R. D.},
  {Randich, S.}, {Bragaglia, A.}, {Carraro, G.}, {Costado, M. T.}, {Flaccomio,
  E.}, {Lanzafame, A. C.}, {Lardo, C.}, {Monaco, L.}, {Morbidelli, L.},
  {Smiljanic, R.}, \& {Zaggia, S.}}]{Jackson_2016}
{Jackson, R. J.}, {Jeffries, R. D.}, {Randich, S.}, {et~al.} 2016, A\&A, 586,
  A52, \dodoi{10.1051/0004-6361/201527507}

\bibitem[{{Jaehnig} {et~al.}(2019){Jaehnig}, {Somers}, \&
  {Stassun}}]{Jaehnig2019}
{Jaehnig}, K., {Somers}, G., \& {Stassun}, K.~G. 2019, \apj, 879, 39,
  \dodoi{10.3847/1538-4357/ab21cf}

\bibitem[{Jayasinghe {et~al.}(2018)Jayasinghe, Kochanek, Stanek, Shappee,
  Holoien, Thompson, Prieto, Dong, Pawlak, Shields, Pojmanski, Otero, Britt, \&
  Will}]{Jayasinghe_2018}
Jayasinghe, T., Kochanek, C.~S., Stanek, K.~Z., {et~al.} 2018, Monthly Notices
  of the Royal Astronomical Society, 477, 3145, \dodoi{10.1093/mnras/sty838}

\bibitem[{Jayasinghe {et~al.}(2019)Jayasinghe, Stanek, Kochanek, Shappee,
  Holoien, Thompson, Prieto, Dong, Pawlak, Pejcha, Shields, Pojmanski, Otero,
  Britt, \& Will}]{Jayasinghe_2019}
Jayasinghe, T., Stanek, K.~Z., Kochanek, C.~S., {et~al.} 2019, Monthly Notices
  of the Royal Astronomical Society, 486, 1907, \dodoi{10.1093/mnras/stz844}

\bibitem[{Jayawardhana {et~al.}(2006)Jayawardhana, Coffey, Scholz, Brandeker,
  \& van Kerkwijk}]{Jayawardhana_2006}
Jayawardhana, R., Coffey, J., Scholz, A., Brandeker, A., \& van Kerkwijk, M.~H.
  2006, The Astrophysical Journal, 648, 1206, \dodoi{10.1086/506171}

\bibitem[{Karim {et~al.}(2016)Karim, Stassun, Brice{\~{n}}o, Vivas, Raetz,
  Mateu, Downes, Calvet, Hern{\'{a}}ndez, Neuh{\"{a}}user, Mugrauer, Takahashi,
  Tachihara, Chini, Cruz-Dias, Aarnio, James, \& Hackstein}]{Karim2016}
Karim, M.~T., Stassun, K.~G., Brice{\~{n}}o, C., {et~al.} 2016, Astron. J.,
  152, 198, \dodoi{10.3847/0004-6256/152/6/198}

\bibitem[{Kawaler(1987)}]{Kawaler_1987}
Kawaler, S.~D. 1987, Publications of the Astronomical Society of the Pacific,
  99, 1322, \dodoi{10.1086/132120}

\bibitem[{Kawaler(1988)}]{Kawaler1988}
---. 1988, Astrophys. J., 333, 236, \dodoi{10.1086/166740}

\bibitem[{{Kesseli} {et~al.}(2018){Kesseli}, {Muirhead}, {Mann}, \&
  {Mace}}]{Kesseli2018}
{Kesseli}, A.~Y., {Muirhead}, P.~S., {Mann}, A.~W., \& {Mace}, G. 2018, \aj,
  155, 225, \dodoi{10.3847/1538-3881/aabccb}

\bibitem[{Kounkel {et~al.}(2017)Kounkel, Hartmann, Loinard, Ortiz-Le{\'{o}}n,
  Mioduszewski, Rodr{\'{\i}}guez, Dzib, Torres, Pech, Galli, Rivera, Boden, II,
  Brice{\~{n}}o, \& Tobin}]{Kounkel_2017}
Kounkel, M., Hartmann, L., Loinard, L., {et~al.} 2017, The Astrophysical
  Journal, 834, 142, \dodoi{10.3847/1538-4357/834/2/142}

\bibitem[{Kounkel {et~al.}(2018)Kounkel, Covey, Su{\'{a}}rez,
  Rom{\'{a}}n-Z{\'{u}}{\~{n}}iga, Hernandez, Stassun, Jaehnig, Feigelson,
  Ram{\'{\i}}rez, Roman-Lopes, Rio, Stringfellow, Kim, Borissova,
  Fern{\'{a}}ndez-Trincado, Burgasser, Garc{\'{\i}}a-Hern{\'{a}}ndez, Zamora,
  Pan, \& Nitschelm}]{Kounkel_2018}
Kounkel, M., Covey, K., Su{\'{a}}rez, G., {et~al.} 2018, The Astronomical
  Journal, 156, 84, \dodoi{10.3847/1538-3881/aad1f1}

\bibitem[{{Kounkel} {et~al.}(2019){Kounkel}, {Covey}, {Moe}, {Kratter},
  {Su{\'a}rez}, {Stassun}, {Rom{\'a}n-Z{\'u}{\~n}iga}, {Hernand ez}, {Kim},
  {Pe{\~n}a Ram{\'\i}rez}, {Roman-Lopes}, {Stringfellow}, {Jaehnig},
  {Borissova}, {Tofflemire}, {Krolikowski}, {Rizzuto}, {Kraus}, {Badenes},
  {Longa-Pe{\~n}a}, {G{\'o}mez Maqueo Chew}, {Barba}, {Nidever}, {Brown}, {De
  Lee}, {Pan}, {Bizyaev}, {Oravetz}, \& {Oravetz}}]{Kounkel_2019}
{Kounkel}, M., {Covey}, K., {Moe}, M., {et~al.} 2019, \aj, 157, 196,
  \dodoi{10.3847/1538-3881/ab13b1}

\bibitem[{{Kuhn} {et~al.}(2019){Kuhn}, {Hillenbrand}, {Sills}, {Feigelson}, \&
  {Getman}}]{Kuhn_2019}
{Kuhn}, M.~A., {Hillenbrand}, L.~A., {Sills}, A., {Feigelson}, E.~D., \&
  {Getman}, K.~V. 2019, \apj, 870, 32, \dodoi{10.3847/1538-4357/aaef8c}

\bibitem[{{Landin} {et~al.}(2016){Landin}, {Mendes}, {Vaz}, \&
  {Alencar}}]{Landin_2016}
{Landin}, N.~R., {Mendes}, L. T.~S., {Vaz}, L. P.~R., \& {Alencar}, S. H.~P.
  2016, A\&A, 586, A96, \dodoi{10.1051/0004-6361/201525851}

\bibitem[{{Lanzafame} {et~al.}(2017){Lanzafame}, {Spada}, \&
  {Distefano}}]{Lanzafame2017}
{Lanzafame}, A.~C., {Spada}, F., \& {Distefano}, E. 2017, \aap, 597, A63,
  \dodoi{10.1051/0004-6361/201628833}

\bibitem[{{Lindegren} {et~al.}(2020{\natexlab{a}}){Lindegren}, {Klioner},
  {Hern{\'a}ndez}, {Bombrun}, {Ramos-Lerate}, {Steidelm{\"u}ller}, {Bastian},
  {Biermann}, {de Torres}, {Gerlach}, {Geyer}, {Hilger}, {Hobbs}, {Lammers},
  {McMillan}, {Stephenson}, {Casta{\~n}eda}, {Davidson}, {Fabricius},
  {Gracia-Abril}, {Portell}, {Rowell}, {Teyssier}, {Torra}, {Bartolom{\'e}},
  {Clotet}, {Garralda}, {Gonz{\'a}lez-Vidal}, {Torra}, {Abbas}, {Altmann},
  {Anglada Varela}, {Balaguer-N{\'u}{\~n}ez}, {Balog}, {Barache}, {Becciani},
  {Bernet}, {Bertone}, {Bianchi}, {Bouquillon}, {Brown}, {Bucciarelli},
  {Busonero}, {Butkevich}, {Buzzi}, {Cancelliere}, {Carlucci}, {Charlot},
  {Cioni}, {Crosta}, {Crowley}, {del Peloso}, {del Pozo}, {Drimmel}, {Esquej},
  {Fienga}, {Fraile}, {Gai}, {Garcia-Reinaldos}, {Guerra}, {Hambly}, {Hauser},
  {Jan{\ss}en}, {Jordan}, {Kostrzewa-Rutkowska}, {Lattanzi}, {Liao}, {Licata},
  {Lister}, {L{\"o}ffler}, {Marchant}, {Masip}, {Mignard}, {Mints}, {Molina},
  {Mora}, {Morbidelli}, {Murphy}, {Pagani}, {Panuzzo}, {Pe{\~n}alosa Esteller},
  {Poggio}, {Re Fiorentin}, {Riva}, {Sagrist{\`a} Sell{\'e}s}, {Sanchez
  Gimenez}, {Sarasso}, {Sciacca}, {Siddiqui}, {Smart}, {Souami}, {Spagna},
  {Steele}, {Taris}, {Utrilla}, {van Reeven}, \& {Vecchiato}}]{Lindegreen2020a}
{Lindegren}, L., {Klioner}, S.~A., {Hern{\'a}ndez}, J., {et~al.}
  2020{\natexlab{a}}, arXiv e-prints, arXiv:2012.03380.
\newblock \doarXiv{2012.03380}

\bibitem[{{Lindegren} {et~al.}(2020{\natexlab{b}}){Lindegren}, {Bastian},
  {Biermann}, {Bombrun}, {de Torres}, {Gerlach}, {Geyer}, {Hern{\'a}ndez},
  {Hilger}, {Hobbs}, {Klioner}, {Lammers}, {McMillan}, {Ramos-Lerate},
  {Steidelm{\"u}ller}, {Stephenson}, \& {van Leeuwen}}]{Lindegreen2020b}
{Lindegren}, L., {Bastian}, U., {Biermann}, M., {et~al.} 2020{\natexlab{b}},
  arXiv e-prints, arXiv:2012.01742.
\newblock \doarXiv{2012.01742}

\bibitem[{{Littlefair}(2014)}]{littlefair2013}
{Littlefair}, S.~P. 2014, in Magnetic Fields throughout Stellar Evolution, ed.
  P.~{Petit}, M.~{Jardine}, \& H.~C. {Spruit}, Vol. 302, 91--99,
  \dodoi{10.1017/S1743921314001793}

\bibitem[{{Lomb}(1976)}]{Lomb_1976}
{Lomb}, N.~R. 1976, \apss, 39, 447, \dodoi{10.1007/BF00648343}

\bibitem[{{Luhman} \& {Esplin}(2020)}]{Luhman2020}
{Luhman}, K.~L., \& {Esplin}, T.~L. 2020, \aj, 160, 44,
  \dodoi{10.3847/1538-3881/ab9599}

\bibitem[{{Majewski} {et~al.}(2017){Majewski}, {Schiavon}, {Frinchaboy},
  {Allende Prieto}, {Barkhouser}, {Bizyaev}, {Blank}, {Brunner}, {Burton},
  {Carrera}, {Chojnowski}, {Cunha}, {Epstein}, {Fitzgerald}, {Garc{\'\i}a
  P{\'e}rez}, {Hearty}, {Henderson}, {Holtzman}, {Johnson}, {Lam}, {Lawler},
  {Maseman}, {M{\'e}sz{\'a}ros}, {Nelson}, {Nguyen}, {Nidever}, {Pinsonneault},
  {Shetrone}, {Smee}, {Smith}, {Stolberg}, {Skrutskie}, {Walker}, {Wilson},
  {Zasowski}, {Anders}, {Basu}, {Beland}, {Blanton}, {Bovy}, {Brownstein},
  {Carlberg}, {Chaplin}, {Chiappini}, {Eisenstein}, {Elsworth}, {Feuillet},
  {Fleming}, {Galbraith-Frew}, {Garc{\'\i}a}, {Garc{\'\i}a-Hern{\'a}ndez},
  {Gillespie}, {Girardi}, {Gunn}, {Hasselquist}, {Hayden}, {Hekker}, {Ivans},
  {Kinemuchi}, {Klaene}, {Mahadevan}, {Mathur}, {Mosser}, {Muna}, {Munn},
  {Nichol}, {O'Connell}, {Parejko}, {Robin}, {Rocha-Pinto}, {Schultheis},
  {Serenelli}, {Shane}, {Silva Aguirre}, {Sobeck}, {Thompson}, {Troup},
  {Weinberg}, \& {Zamora}}]{Majewski2017}
{Majewski}, S.~R., {Schiavon}, R.~P., {Frinchaboy}, P.~M., {et~al.} 2017, \aj,
  154, 94, \dodoi{10.3847/1538-3881/aa784d}

\bibitem[{{Makidon} {et~al.}(2004){Makidon}, {Rebull}, {Strom}, {Adams}, \&
  {Patten}}]{Makidon2004}
{Makidon}, R.~B., {Rebull}, L.~M., {Strom}, S.~E., {Adams}, M.~T., \& {Patten},
  B.~M. 2004, \aj, 127, 2228, \dodoi{10.1086/382237}

\bibitem[{{Manzo-Mart{\'\i}nez} {et~al.}(2020){Manzo-Mart{\'\i}nez}, {Calvet},
  {Hern{\'a}ndez}, {Lizano}, {Hern{\'a}ndez}, {Miller}, {Mauc{\'o}},
  {Brice{\~n}o}, \& {D'Alessio}}]{Manzo_2020}
{Manzo-Mart{\'\i}nez}, E., {Calvet}, N., {Hern{\'a}ndez}, J., {et~al.} 2020,
  \apj, 893, 56, \dodoi{10.3847/1538-4357/ab7ead}

\bibitem[{Marigo {et~al.}(2017)Marigo, Girardi, Bressan, Rosenfield, Aringer,
  Chen, Dussin, Nanni, Pastorelli, Rodrigues, Trabucchi, Bladh, Dalcanton,
  Groenewegen, Montalb{\'{a}}n, \& Wood}]{Marigo2017}
Marigo, P., Girardi, L., Bressan, A., {et~al.} 2017, Astrophys. J., 835, 77,
  \dodoi{10.3847/1538-4357/835/1/77}

\bibitem[{{Marilli, E.} {et~al.}(2007){Marilli, E.}, {Frasca, A.}, {Covino,
  E.}, {Alcal\'a, J. M.}, {Catalano, S.}, {Fern\'andez, M.}, {Arellano Ferro,
  A.}, {Rubio-Herrera, E.}, \& {Spezzi, L.}}]{Marilli_2007}
{Marilli, E.}, {Frasca, A.}, {Covino, E.}, {et~al.} 2007, A\&A, 463, 1081,
  \dodoi{10.1051/0004-6361:20066458}

\bibitem[{{Matt} {et~al.}(2012){Matt}, {Pinz{\'o}n}, {Greene}, \&
  {Pudritz}}]{Matt2012}
{Matt}, S.~P., {Pinz{\'o}n}, G., {Greene}, T.~P., \& {Pudritz}, R.~E. 2012,
  \apj, 745, 101, \dodoi{10.1088/0004-637X/745/1/101}

\bibitem[{{Maxted} {et~al.}(2008){Maxted}, {Jeffries}, {Oliveira}, {Naylor}, \&
  {Jackson}}]{Maxted2008}
{Maxted}, P.~F.~L., {Jeffries}, R.~D., {Oliveira}, J.~M., {Naylor}, T., \&
  {Jackson}, R.~J. 2008, \mnras, 385, 2210,
  \dodoi{10.1111/j.1365-2966.2008.13008.x}

\bibitem[{Morales-Calder{\'{o}}n {et~al.}(2011)Morales-Calder{\'{o}}n,
  Stauffer, Hillenbrand, Gutermuth, Song, Rebull, Plavchan, Carpenter, Whitney,
  Covey, de~Oliveira, Winston, McCaughrean, Bouvier, Guieu, Vrba, Holtzman,
  Marchis, Hora, Wasserman, Terebey, Megeath, Guinan, Forbrich, Hu{\'{e}}lamo,
  Riviere-Marichalar, Barrado, Stapelfeldt, Hern{\'{a}}ndez, Allen, Ardila,
  Bayo, Favata, James, Werner, \& Wood}]{Morales_Calderon_2011}
Morales-Calder{\'{o}}n, M., Stauffer, J.~R., Hillenbrand, L.~A., {et~al.} 2011,
  The Astrophysical Journal, 733, 50, \dodoi{10.1088/0004-637x/733/1/50}

\bibitem[{{Murdin} \& {Penston}(1977)}]{Murdin_1977}
{Murdin}, P., \& {Penston}, M.~V. 1977, \mnras, 181, 657,
  \dodoi{10.1093/mnras/181.4.657}

\bibitem[{Newville {et~al.}(2014)Newville, Stensitzki, Allen, \&
  Ingargiola}]{Newville_2014}
Newville, M., Stensitzki, T., Allen, D.~B., \& Ingargiola, A. 2014, {LMFIT:
  Non-Linear Least-Square Minimization and Curve-Fitting for Python}, 0.8.0,
  Zenodo, \dodoi{10.5281/zenodo.11813}

\bibitem[{Nguyen {et~al.}(2009)Nguyen, Jayawardhana, van Kerkwijk, Brandeker,
  Scholz, \& Damjanov}]{Nguyen_2009}
Nguyen, D.~C., Jayawardhana, R., van Kerkwijk, M.~H., {et~al.} 2009, The
  Astrophysical Journal, 695, 1648, \dodoi{10.1088/0004-637x/695/2/1648}

\bibitem[{Nicholson {et~al.}(2018)Nicholson, Hussain, Donati, Folsom, Mengel,
  Carter, Wright, \& collaboration}]{Nicholson_2018}
Nicholson, B.~A., Hussain, G. A.~J., Donati, J.-F., {et~al.} 2018, Monthly
  Notices of the Royal Astronomical Society, 480, 1754,
  \dodoi{10.1093/mnras/sty1965}

\bibitem[{{Olney} {et~al.}(2020){Olney}, {Kounkel}, {Schillinger}, {Scoggins},
  {Yin}, {Howard}, {Covey}, {Hutchinson}, \& {Stassun}}]{Olney_2020}
{Olney}, R., {Kounkel}, M., {Schillinger}, C., {et~al.} 2020, \aj, 159, 182,
  \dodoi{10.3847/1538-3881/ab7a97}

\bibitem[{Parihar {et~al.}(2009)Parihar, Messina, Distefano, Shantikumar, \&
  Medhi}]{Parihar_2009}
Parihar, P., Messina, S., Distefano, E., Shantikumar, N.~S., \& Medhi, B.~J.
  2009, Monthly Notices of the Royal Astronomical Society, 400, 603,
  \dodoi{10.1111/j.1365-2966.2009.15496.x}

\bibitem[{{Pecaut} \& {Mamajek}(2013)}]{Pecaut2013}
{Pecaut}, M.~J., \& {Mamajek}, E.~E. 2013, \apjs, 208, 9,
  \dodoi{10.1088/0067-0049/208/1/9}

\bibitem[{Pinz{\'{o}}n {et~al.}(2021)Pinz{\'{o}}n, Hern{\'{a}}ndez, Serna,
  Garc{\'{\i}}a, Manzo-Mart{\'{\i}}nez, Roman-Lopes,
  Rom{\'{a}}n-Z{\'{u}}{\~{n}}iga, Batista, Ram{\'{\i}}rez-V{\'{e}}lez, Osorio,
  \& Avenda{\~{n}}o}]{Pinzon2021}
Pinz{\'{o}}n, G., Hern{\'{a}}ndez, J., Serna, J., {et~al.} 2021, The
  Astronomical Journal, 162, 90, \dodoi{10.3847/1538-3881/ac04ae}

\bibitem[{Rebull(2001)}]{Rebull2001}
Rebull, L.~M. 2001, The Astronomical Journal, 121, 1676, \dodoi{10.1086/319393}

\bibitem[{Rebull {et~al.}(2006)Rebull, Stauffer, Megeath, Hora, \&
  Hartmann}]{Rebull2006a}
Rebull, L.~M., Stauffer, J.~R., Megeath, S.~T., Hora, J.~L., \& Hartmann, L.
  2006, Astrophys. J., 646, 297, \dodoi{10.1086/504865}

\bibitem[{Rebull {et~al.}(2014)Rebull, Cody, Covey, Günther, Hillenbrand,
  Plavchan, Poppenhaeger, Stauffer, Wolk, Gutermuth, Morales-Calder{\'{o}}n,
  Song, Barrado, Bayo, James, Hora, Vrba, de~Oliveira, Bouvier, Carey,
  Carpenter, Favata, Flaherty, Forbrich, Hernandez, McCaughrean, Megeath,
  Micela, Smith, Terebey, Turner, Allen, Ardila, Bouy, \& Guieu}]{Rebull_2014}
Rebull, L.~M., Cody, A.~M., Covey, K.~R., {et~al.} 2014, The Astronomical
  Journal, 148, 92, \dodoi{10.1088/0004-6256/148/5/92}

\bibitem[{Reiners \& Schmitt(2002)}]{Reiners2002}
Reiners, A., \& Schmitt, J. H. M.~M. 2002, Astron. Astrophys., 384, 155,
  \dodoi{10.1051/0004-6361:20011801}

\bibitem[{Rhode {et~al.}(2001)Rhode, Herbst, \& Mathieu}]{Rhode2001}
Rhode, K.~L., Herbst, W., \& Mathieu, R.~D. 2001, Astron. J., 122, 3258,
  \dodoi{10.1086/324448}

\bibitem[{Ricker {et~al.}(2014)Ricker, Winn, Vanderspek, Latham, Bakos, Bean,
  Berta-Thompson, Brown, Buchhave, Butler, Butler, Chaplin, Charbonneau,
  Christensen-Dalsgaard, Clampin, Deming, Doty, Lee, Dressing, Dunham, Endl,
  Fressin, Ge, Henning, Holman, Howard, Ida, Jenkins, Jernigan, Johnson,
  Kaltenegger, Kawai, Kjeldsen, Laughlin, Levine, Lin, Lissauer, MacQueen,
  Marcy, McCullough, Morton, Narita, Paegert, Palle, Pepe, Pepper, Quirrenbach,
  Rinehart, Sasselov, Sato, Seager, Sozzetti, Stassun, Sullivan, Szentgyorgyi,
  Torres, Udry, \& Villasenor}]{Ricker2014}
Ricker, G.~R., Winn, J.~N., Vanderspek, R., {et~al.} 2014, 9143, 556 ,
  \dodoi{10.1117/12.2063489}

\bibitem[{{Royer}(2005)}]{Royer2005}
{Royer}, F. 2005, Memorie della Societa Astronomica Italiana Supplementi, 8,
  124

\bibitem[{{Royer} {et~al.}(2002){Royer}, {Gerbaldi}, {Faraggiana}, \&
  {G{\'o}mez}}]{Royer_2002}
{Royer}, F., {Gerbaldi}, M., {Faraggiana}, R., \& {G{\'o}mez}, A.~E. 2002,
  \aap, 381, 105, \dodoi{10.1051/0004-6361:20011422}

\bibitem[{{Sacco} {et~al.}(2008){Sacco}, {Franciosini}, {Randich}, \&
  {Pallavicini}}]{Sacco_2008}
{Sacco}, G.~G., {Franciosini}, E., {Randich}, S., \& {Pallavicini}, R. 2008,
  A\&A, 488, 167, \dodoi{10.1051/0004-6361:20079049}

\bibitem[{{Sauty} {et~al.}(2011){Sauty}, {Meliani}, {Lima}, {Tsinganos},
  {Cayatte}, \& {Globus}}]{Sauty2011}
{Sauty}, C., {Meliani}, Z., {Lima}, J. J.~G., {et~al.} 2011, A\&A, 533, A46,
  \dodoi{10.1051/0004-6361/201116519}

\bibitem[{{Scargle}(1982)}]{Scargle_1982}
{Scargle}, J.~D. 1982, \apj, 263, 835, \dodoi{10.1086/160554}

\bibitem[{Sicilia-Aguilar {et~al.}(2005)Sicilia-Aguilar, Hartmann,
  Szentgyorgyi, Fabricant, F{\H{u}}r{\'{e}}sz, Roll, Conroy, Calvet, Tokarz, \&
  Hern{\'{a}}ndez}]{Sicilia_Aguilar_2005}
Sicilia-Aguilar, A., Hartmann, L.~W., Szentgyorgyi, A.~H., {et~al.} 2005, The
  Astronomical Journal, 129, 363, \dodoi{10.1086/426327}

\bibitem[{Sim{\'{o}}n-D{\'{i}}az \& Herrero(2007)}]{Simon-Diaz2007}
Sim{\'{o}}n-D{\'{i}}az, S., \& Herrero, A. 2007, Astron. Astrophys., 468, 1063,
  \dodoi{10.1051/0004-6361:20066060}

\bibitem[{Skumanich(1972)}]{Skumanich1972}
Skumanich, A. 1972, Astrophys. J., 171, 565, \dodoi{10.1086/151310}

\bibitem[{{Somers} {et~al.}(2020){Somers}, {Cao}, \&
  {Pinsonneault}}]{Somers2020}
{Somers}, G., {Cao}, L., \& {Pinsonneault}, M.~H. 2020, \apj, 891, 29,
  \dodoi{10.3847/1538-4357/ab722e}

\bibitem[{{Somers} \& {Pinsonneault}(2015)}]{Somers2015}
{Somers}, G., \& {Pinsonneault}, M.~H. 2015, \apj, 807, 174,
  \dodoi{10.1088/0004-637X/807/2/174}

\bibitem[{{Somers} \& {Stassun}(2017)}]{Somers2017}
{Somers}, G., \& {Stassun}, K.~G. 2017, \aj, 153, 101,
  \dodoi{10.3847/1538-3881/153/3/101}

\bibitem[{{Soubiran} {et~al.}(2019){Soubiran}, {Cantat-Gaudin},
  {Romero-G{\'o}mez}, {Casamiquela}, {Jordi}, {Vallenari}, {Antoja},
  {Balaguer-N{\'u}{\~n}ez}, {Bossini}, {Bragaglia}, {Carrera}, {Castro-Ginard},
  {Figueras}, {Heiter}, {Katz}, {Krone-Martins}, {Le Campion}, {Moitinho}, \&
  {Sordo}}]{Soubiran_2019}
{Soubiran}, C., {Cantat-Gaudin}, T., {Romero-G{\'o}mez}, M., {et~al.} 2019,
  \aap, 623, C2, \dodoi{10.1051/0004-6361/201834020e}

\bibitem[{{Spada} \& {Lanzafame}(2020)}]{Spada_2020}
{Spada}, F., \& {Lanzafame}, A.~C. 2020, A\&A, 636, A76,
  \dodoi{10.1051/0004-6361/201936384}

\bibitem[{{Stassun}(2019)}]{Stassun2019}
{Stassun}, K.~G. 2019, VizieR Online Data Catalog, IV/38

\bibitem[{Stassun {et~al.}(1999)Stassun, Mathieu, Mazeh, \& Vrba}]{Stassun1999}
Stassun, K.~G., Mathieu, R.~D., Mazeh, T., \& Vrba, F.~J. 1999, Astron. J.,
  117, 2941, \dodoi{10.1086/300881}

\bibitem[{{Still} \& {Barclay}(2012)}]{PyKE}
{Still}, M., \& {Barclay}, T. 2012, {PyKE: Reduction and analysis of Kepler
  Simple Aperture Photometry data}.
\newblock \doeprint{1208.004}

\bibitem[{{Testi} {et~al.}(2014){Testi}, {Birnstiel}, {Ricci}, {Andrews},
  {Blum}, {Carpenter}, {Dominik}, {Isella}, {Natta}, {Williams}, \&
  {Wilner}}]{Testi2014}
{Testi}, L., {Birnstiel}, T., {Ricci}, L., {et~al.} 2014, in Protostars and
  Planets VI, ed. H.~{Beuther}, R.~S. {Klessen}, C.~P. {Dullemond}, \&
  T.~{Henning}, 339, \dodoi{10.2458/azu_uapress_9780816531240-ch015}

\bibitem[{{Thanathibodee} {et~al.}(2020){Thanathibodee}, {Molina}, {Calvet},
  {Serna}, {Bae}, {Reynolds}, {Hern{\'a}ndez}, {Muzerolle}, \&
  {Hern{\'a}ndez}}]{Thanathibodee_2020}
{Thanathibodee}, T., {Molina}, B., {Calvet}, N., {et~al.} 2020, \apj, 892, 81,
  \dodoi{10.3847/1538-4357/ab77c1}

\bibitem[{Venuti {et~al.}(2017)Venuti, Bouvier, Cody, Stauffer, Micela, Rebull,
  Alencar, Sousa, Hillenbrand, \& Flaccomio}]{Venuti2017}
Venuti, L., Bouvier, J., Cody, A.~M., {et~al.} 2017, Astron. Astrophys., 599,
  A23, \dodoi{10.1051/0004-6361/201629537}

\bibitem[{{Warren} \& {Hesser}(1977)}]{Warren_1977}
{Warren}, W.~H., J., \& {Hesser}, J.~E. 1977, \apjs, 34, 115,
  \dodoi{10.1086/190446}

\bibitem[{{Williams} \& {Cieza}(2011)}]{Williams2011}
{Williams}, J.~P., \& {Cieza}, L.~A. 2011, \araa, 49, 67,
  \dodoi{10.1146/annurev-astro-081710-102548}

\bibitem[{Wolff {et~al.}(2004)Wolff, Strom, \& Hillenbrand}]{Wolff_2004}
Wolff, S.~C., Strom, S.~E., \& Hillenbrand, L.~A. 2004, The Astrophysical
  Journal, 601, 979, \dodoi{10.1086/380503}

\bibitem[{{Zari} {et~al.}(2019){Zari}, {Brown}, \& {de Zeeuw}}]{Zari2019}
{Zari}, E., {Brown}, A.~G.~A., \& {de Zeeuw}, P.~T. 2019, \aap, 628, A123,
  \dodoi{10.1051/0004-6361/201935781}

\end{thebibliography}
\bibliographystyle{aasjournal}

\appendix
\section{Rotational velocities and rotational periods for stars not included as kinematic or spectroscopic members} \label{apendiceA}
\begin{longrotatetable}
\begin{deluxetable*}{cccccccccccc}
\tablecaption{Rotational velocities and rotational periods for stars not included as kinematic or spectroscopic members \label{tab:apendiceA}}
\tablehead{
\colhead{2massID}   &  \colhead{$v\sin(i)$\tablenotemark{a}} & \colhead{$v\sin(i)$\tablenotemark{b}}  & \colhead{Binary\tablenotemark{a}} &\colhead{TIC} & \colhead{Tmag} &\colhead{Period\tablenotemark{c}} & \colhead{Period\tablenotemark{d}} &\colhead{References\tablenotemark{d}} \\
\colhead{       }   &  \colhead{km/s}                           & \colhead{km/s}                      & \colhead{      }     &\colhead{ } & \colhead{mag} &\colhead{days} & \colhead{days} &\colhead{}
}
\startdata
  05015108-0436117   &  \nodata  &  \nodata  &	\nodata	&    213048173   &    12.07   &     5.50 $\pm$  0.01 	 &  \nodata  &  \nodata \\
  05521912-0557431   &  \nodata  &  \nodata  &	\nodata	&     66887850   &    12.34   &     1.95 $\pm$  0.01 	 &  \nodata  &  \nodata \\
  05522239+0017292   &      2.9 $\pm$  14.7   &  \nodata  &	1	&  \nodata  &  \nodata  &  \nodata   &  \nodata  &  \nodata \\
  05522245-0012264   &      0.1 $\pm$   1.0   &  \nodata  &	1	&  \nodata  &  \nodata  &  \nodata   &  \nodata  &  \nodata \\
  05522363+0007020   &     35.2 $\pm$   0.1   &  \nodata  &	2	&  \nodata  &  \nodata  &  \nodata   &  \nodata  &  \nodata \\
  05523827-0657042   &  \nodata  &  \nodata  &	\nodata	&     66888753   &    13.11   &     2.80 $\pm$  0.01 	 &  \nodata  &  \nodata \\
  05524880-0644372   &  \nodata  &  \nodata  &	\nodata	&     66888587   &    13.66   &    11.91 $\pm$  0.04 	 &  \nodata  &  \nodata \\
  05530076-0747254   &  \nodata  &  \nodata  &	\nodata	&     66913082   &    14.88   &     9.63 $\pm$  0.10 	 &  \nodata  &  \nodata \\
  05535213+0133074   &  \nodata  &  \nodata  &	\nodata	&    159089283   &    14.14   &     5.74 $\pm$  0.03 	 &  \nodata  &  \nodata \\
\enddata 
\tablenotetext{a}{\citet{Kounkel_2018,Kounkel_2019}: 0- Undeconvolvable cross-correlation function (CCF); 1- Only a single component in the CCF; 2- Multiple components in the CCF; -1- Spotted pairs or SB2 Uncertain}
\tablenotetext{b}{Fourier Method, \S\ref{sec:vsini}}
\tablenotetext{c}{TESS periods \S\ref{sec:tessperiods}}
\tablenotetext{d}{Known periods:1) \citet{Stassun1999}; 2)\citet{Rebull2001,Rebull2006a}; 3) \citet[][;J-band]{Carpenter2001}; 4) \citet{Herbst_2002}; 5) \citet{Cody2010}; 6) \citet{Morales_Calderon_2011}; 7) \citet[7]{Karim2016}}
\tablecomments{Only a portion of the table is shown here. The full version is available in electronic form.}
\end{deluxetable*}
\end{longrotatetable}

\end{document}